\documentclass[journal]{new-aiaa}

\usepackage[monochrome]{xcolor}

\usepackage{algorithm,algorithmic}
\usepackage{subfigure}
\usepackage{psfrag}
\usepackage{bm}
\usepackage{graphicx}
\usepackage{amsmath}
\usepackage{amsfonts}
\usepackage[version=4]{mhchem}
\usepackage{siunitx}
\usepackage{longtable,tabularx}
\begin{document}

\title{Dynamics Near the Three-Body Libration Points \\ via Koopman Operator Theory}

\author{Simone Servadio \footnote{Postdoctoral Associate, Department of Aeronautics and Astronautics, simoserv@mit.edu. (Corresponding Author)}}
\affil{Massachusetts Institute of Technology, Cambridge, MA 02139, USA}
\author{David Arnas\footnote{Assistant Professor, School of Aeronautics and Astronautics, Purdue University, darnas@purdue.edu.}}
\affil{Purdue University, West Lafayette, IN 47907, USA}
\author{Richard Linares\footnote{Boeing Assistant Professor, Department of Aeronautics and Astronautics, Email: linaresr@mit.edu, Senior Member AIAA.}}
\affil{Massachusetts Institute of Technology, Cambridge, MA 02139, USA}

 \maketitle
 
\begin{abstract}
This paper investigates the application of the Koopman Operator theory to the motion of a satellite about a libration point in the Circular Restricted Three-Body Problem. Recently, the Koopman Operator has emerged as a promising alternative to the geometric perspective for dynamical systems, where the Koopman Operator formulates the analysis and dynamical systems in terms of observables. This paper explores the use of the Koopman Operator for computing both 2D and 3D periodic orbits near libration points. Further, simulation results show that the Koopman Operator provides analytical solutions with high accuracy for both Lyapunov and Halo orbits, which are then applied to a station-keeping application.
\end{abstract}

%%%%%%%%%%%%%%%%%%%%%%%%%%%%%%%%%%%%%%%%%%%%%%%%%%%%%%%%%%%%%%%%%%%%%%%%%%%%%%%%%%%%%%%%%%%%%%
% Introduction
%%%%%%%%%%%%%%%%%%%%%%%%%%%%%%%%%%%%%%%%%%%%%%%%%%%%%%%%%%%%%%%%%%%%%%%%%%%%%%%%%%%%%%%%%%%%%%
 
\section{Introduction}

Periodic orbits in the circular Restricted Three-Body Problem (RTBP), such as Lissajous and Halo type trajectories, are of high interest for space mission design applications. In particular, many missions, such as the cislunar space gateway concept, have been proposed that utilize these orbits. The proposed orbit for the gateway is a Near-Rectilinear Halo Orbit (NRHO), which is a non-Keplerian trajectory with the favorable properties of a continuous line of sight coverage for communications with Earth and fuel-efficient access to the lunar surface~\cite{whitley2016options}. Although operating in the RTBP region while utilizing non-Keplerian trajectories has its benefits, it remains challenging to develop, analyze, and perform guidance, navigation, and control for these missions due to the nonlinearities of the RTBP region. Therefore, new approaches for the analytical analysis of these missions are needed.

Analytical approaches exist for analyzing the RTBP and developing solutions for their trajectories. In particular, Richardson~\cite{richardson1980analytic} developed a Lindstedt–Poincaré procedure for computing the Halo orbits through matching the $x$-$y$ periodic frequency with that of the $z$ direction motion. Lindstedt–Poincaré methods are powerful perturbation approaches, but they require extensive algebraic computation and can be difficult to derive for higher-order solutions. In addition to Lindstedt–Poincaré method, Lie-Deprit methods have been successfully applied to the central gravity and RTBP problems~\cite{elipe1987rotation}, with extensive applications, from reduction center manifold~\cite{jorba1999dynamics} to long term propagation~\cite{lara2008computing}. Particularly, the evaluation of periodic orbits around the libration points is quite efficient when approached in complex variables~\cite{lara2018higher}.

A general framework for the analysis, estimation, and control of nonlinear systems remains an engineering grand challenge~\cite{junkins2004nonlinear}. Dealing with nonlinear dynamics and measurement models has been an active area of research~\cite{terejanu2008uncertainty,horwood2011gaussian,madankan2013polynomial}. A promising direction for the analysis of nonlinear systems is to seek a transformation that embeds the nonlinear dynamics in a global linear representation~\cite{hurtadousing}. The Koopman Operator (KO)~\cite{koopman1931hamiltonian} provides such a framework and has been applied to many high dimensional and nonlinear problems in engineering~\cite{mezic2013analysis,
mezic2004comparison,mezic2020koopman,budivsic2012applied,mezic2005spectral,williams2015data}, as well as to the zonal harmonics problem in astrodynamics~\cite{zonal_koopman}. This paper focuses instead on the application of the KO theory to obtain analytical orbit methodologies for the circular Restricted Three-Body Problem. 

The goal of the proposed KO-based method is reformulating nonlinear problems that arise in astrodynamics into a linear framework that can be solved with available linear techniques. Motivated by Koopman~\cite{koopman1931hamiltonian}  and Von Neumann~\cite{neumann1932operatorenmethode}, a new  ``Heisenberg picture” was introduced in classical dynamics. In the work by Koopman, it was observed that a KO for a Hamiltonian system is unitary in an $L_2$ Hilbert space~\cite{koopman1931hamiltonian}. Later Von Neumann~\cite{neumann1932operatorenmethode} was able to make a connection between the spectrum of the KO and that of ergodicity of classical dynamical systems. The key result was demonstrating that there exists an infinite-dimensional linear operator, given by $\mathcal{K}$, that evolves all observation functions $g({\bf x})$ of the state, ${\bf x}$, for any nonlinear system. The evolution of these observables and the KO is defined by the chain rule, which is a linear operator, giving KO its linear properties. 

The linearity of the KO is very appealing, but this benefit is contrasted with the fact that it is infinite-dimensional. However, this issue can be overcome by capturing the evolution on a finite subspace spanned by a finite set of basis functions instead of capturing the evolution of all measurement functions in a Hilbert space~\cite{brunton2016koopman}. In effect, this is a truncation of the KO to a finite subspace.  Additionally, a Koopman invariant subspace is spanned by a set of eigenfunctions of the Koopman operator. A Koopman eigenfunction, $\phi_i({\bf x})$, corresponding to eigenvalue $\lambda_i$ is invariant under the Koopman operator (but for a normalized constant). As such, the evolution of the Koopman eigenfunctions can be expressed as $\frac{d}{dt}\phi_i({\bf x})=\lambda_i\phi_i({\bf x})$. The technique also provides an accurate spectral analysis of partial differential equations (PDEs)~\cite{nakao2020spectral,kutz2016koopman}.

The main contribution of the paper is the development of an analytical solution of the RTBP through the KO. Consequently, the dynamics are linearized and described through well-selected eigenfunctions, which describe the evolution of the state of the system over time. A detailed study of the eigenvalues of the KO matrix gives information regarding the stable and unstable part of the system, with a particular focus on its frequencies. As such, the modes are evaluated and spectral analysis is conducted. The robustness and accuracy of the newly developed technique is assessed through a convergence analysis for different KO orders. Moreover, the paper presents a new particle filtering technique where the propagation of samples is achieved through the Koopman solution and a new model predictive controller that optimizes the feedback control input using the KO to predict, in time, the state of the system.   

The organization of this paper is as follows. First, the KO theory and the Galerkin method used in this paper are summarized. Next, the dynamics near the $\mathcal L_1$, $\mathcal L_2$, and $\mathcal L_3$ points are described along with the complex normal form Hamiltonian for the RTBP model. Then, simulation results are provided for both two-dimensional and three-dimensional periodic orbits, proposing the KO solution for Lyapounov and Halo orbits. The accuracy of the solution is assessed through a convergence analysis. The spectral behavior of the system is studied in terms of its modes and frequencies. A comparison between the Galerkin methodology and the Extended Dynamics Model Decomposition is also performed. Lastly, a station-keeping application uses new developed KO-based filter and controller to keep a spacecraft orbiting around the desired Halo orbit.

%%%%%%%%%%%%%%%%%%%%%%%%%%%%%%%%%%%%%%%%%%%%%%%%%%%%%%%%%%%%%%%%%%%%%%%%%%%%%%%%%%%%%%%%%%%%%%
% Preliminaries
%%%%%%%%%%%%%%%%%%%%%%%%%%%%%%%%%%%%%%%%%%%%%%%%%%%%%%%%%%%%%%%%%%%%%%%%%%%%%%%%%%%%%%%%%%%%%%

\section{Koopman Operator Theory}

A classical definition of nonlinear dynamical systems is given by the initial value problem, which can be represented by a set of coupled autonomous ordinary differential equations in the form:
\begin{equation}\label{problem}
\left\{ \begin{tabular}{l}
    $\displaystyle\frac{d}{dt}{\bf x}(t) = {\bf f}({\bf x})$ \\
    ${\bf x}(t_0) = {\bf x_0}$ 
\end{tabular}  \right.
\end{equation}
where ${\bf x}\in\mathbb{R}^d$ is the state which depends on the time evolution $t$, ${\bf f}: \mathbb{R}^d\rightarrow \mathbb{R}^d$ is the nonlinear dynamics model, $d$ is the number of dimensions in which the problem is defined, and ${\bf x_0}$ is the initial condition of the system at time $t_0$. The KO $(\mathcal{K})$ is an infinite-dimensional linear operator that evolves all observable functions $\mathcal{G}({\bf x})$ of the state, allowing to define any problem of classical mechanics in operator form.

Let $\mathcal{F}$ be a vector space of observable functions, where $\mathcal{G}({\bf x})\in \mathcal{F}$. Since the KO is an infinite-dimensional linear operator, this space of functions $\mathcal{F}$, in which the observables are defined on, is also infinite-dimensional. Therefore, if $g\subseteq\mathcal{G}({\bf x})$ is a given observable in this space, the evolution of $g$ in the dynamical system is represented by:
\begin{equation}
\mathcal{K}\left(g({\bf x})\right) = \frac{d}{dt}g({\bf x}) = \left( \nabla_{{\bf x}} g({\bf x})\right)\frac{d}{dt}{\bf x}(t) = \left( \nabla_{{\bf x}} g({\bf x})\right){\bf f}({\bf x}),
\end{equation}
where $\nabla_{{\bf x}} g = (\partial g/\partial x_1,\partial g/\partial x_2,\dots,\partial g/\partial x_d)$. That way, the evolution of any observable subjected to the dynamical system is provided by the Koopman Operator:
\begin{equation}
\mathcal{K}\left(\cdot\right) = \left( \nabla_{{\bf x}} \cdot\right){\bf f}({\bf x}).
\end{equation}

Note that the evolution of the observables is provided by the application of the chain rule for the time derivative of $g({\bf x})$. Consequently, the defined operator is linear, in that:
\begin{equation}
    \mathcal{K}\left(\beta_1g_1({\bf x})+\beta_2g_2({\bf x})\right)=\beta_1\mathcal{K}\left(g_1({\bf x})\right)+\beta_2\mathcal{K}\left(g_2({\bf x})\right),
\end{equation}
for any pair of observables $g_1\subseteq\mathcal{G}({\bf x})$ and $g_2\subseteq\mathcal{G}({\bf x})$ and any arbitrary constants $\beta_1$ and $\beta_2$. This property was outlined in Koopman's paper~\cite{koopman1931hamiltonian}. The linearity of the Koopman Operator is very appealing, but this benefit is contrasted with the fact that it is infinite-dimensional. However, this issue can be overcome by capturing the evolution of the system on a finite subspace spanned by a finite set of basis functions instead of all measurement functions in a Hilbert space~\cite{brunton2016koopman}. In effect, this is a truncation of the Koopman operator to a finite subspace $\mathcal{F}_D$ of dimension $m$, where $\mathcal{F}_D \in \mathcal{F}$. This subspace $\mathcal{F}_D$ can be spanned by any set of eigenfunctions $\phi_i\in\mathcal{F}_D$, with $i\in\{1,2,\dots,m\}$, defined as:
\begin{equation}\label{koopman}
\mathcal{K}\left(\phi_i({\bf x})\right) =\frac{d}{dt}\phi_i({\bf x})=\lambda_i \phi_i({\bf x}),
\end{equation}
where $\lambda_i$ are the eigenvalues associated with the eigenfunctions $\phi_i$, and $m$ is the number of eigenfunctions chosen to represent the space. Therefore, the Koopman eigenfunctions can be used to form a transformation of variables that linearizes the system. Particularly, let ${\Phi}({\bf x}) = \left(\phi_1({\bf x}), \dots, \phi_m({\bf x}) \right)^T$ be the set of eigenfunctions of the KO in $\mathcal{F}_D$. Then, using the relation in Eq.~\eqref{koopman}, it is possible to write the evolution of ${\bf \Phi}$ as: 
\begin{equation}\label{time_eigenf}
\mathcal{K}\left(\bf \Phi\right) = \frac{d}{dt}{\bf \Phi}=\Lambda {\bf \Phi},
\end{equation}
where $\Lambda=\text{diag}([\lambda_1, \dots,\lambda_m])$ is the diagonal matrix containing the eigenvalues of the system in $\mathcal{F}_D$. This transformation is called the Koopman Canonical Transform~\cite{surana2016linear}. The solution of Eq.~\eqref{time_eigenf} is:
\begin{equation}\label{eigen_time}
    {\bf \Phi}(t) = \exp(\Lambda t){\bf \Phi}(t_0),
\end{equation}
where ${\Phi}(t_0)$ is the value of the eigenfunctions at the initial time $t_0$. This result will be used later to solve the complete system once the eigenfunctions of the operator are obtained.

In general, we are interested in the identity observable, that is, ${\bf g}({\bf x})={\bf x}$. Therefore, it is required to be able to represent these observables in terms of the KO eigenfunctions. This is achieved using the Koopman modes, i. e., the projection of the full-state observable onto the KO eigenfunctions. If this projection can be found, the evolution of the state is represented by means of the evolution of the KO eigenfunctions, and thus, an approximate solution to the system can be provided. However, it is important to note that the challenge resides in computing these eigenfunctions and eigenvalues of the system.

%*****************************************************************************
%*****************************************************************************
%*****************************************************************************

\subsection{Computing the Koopman eigenfunctions via Galerkin Method}
This section discusses the use of the Galerkin method for computing the eigenfunctions of the KO. First, the KO is used to define a Partial Differential Equation (PDE) for the time evolution of a scalar function $u({\bf x},t)$ (note that in general $u({\bf x},t)$ is a function of time, $t$, and ${\bf x}$). The Galerkin method is then used to convert the time evolution PDE to a matrix form using a series expansion for $u({\bf x},t)$ over a predefined basis set. This matrix form can be used to solve for the eigenfunctions and eigenvalues of the KO. 
The Koopman Operator defines a first-order PDE for the time evolution of a scale function $u({\bf x},t)$
\begin{equation}\label{koopman_PDE1}
\frac{d u({\bf x},t)}{dt}
={ f}_1\left({\bf x}\right)\frac{\partial }{\partial x_1} u({\bf x},t) +\cdots+  { f}_d\left({\bf x}\right)\frac{\partial }{\partial x_d}u({\bf x},t),
\end{equation}
Additionally, the eigenfunctions of the Koopman Operator give rise to a set of linear first-order PDEs for the eigenfunctions in the form: 
\begin{equation}\label{koopman2}
\mathcal{K}(\phi_i) = \left(\nabla_x \phi_i({\bf x})\right) {\bf f}({\bf x}) =\lambda_i \phi_i({\bf x}),
\end{equation}
or in a more expanded notation:
\begin{equation}\label{koopman_PDE}
\frac{d \phi_i({\bf x})}{dt}={ f}_1\left({\bf x}\right)\frac{\partial }{\partial x_1} \phi_i({\bf x}) +\cdots+  { f}_d\left({\bf x}\right)\frac{\partial }{\partial x_d}\phi_i({\bf x}) =\lambda_i \phi_i({\bf x}),
\end{equation}
where ${\bf f}({\bf x}) = (f_1({\bf x}), f_2({\bf x}), \dots, f_d({\bf x}))^T$. The above PDE is an advection equation for the ``concentration" function $\phi_i({\bf x})$ with velocity ${\bf f}({\bf x})$. This equation is a linear first-order PDE and in general has no closed-form solution. However, it is possible to approximate the solution of this linear PDE using the Galerkin method. 

This work makes use of normalized Legendre polynomials as proposed by Arnas and Linares~\cite{zonal_koopman}, due to the advantages they provide in the computation of the Koopman matrix. Legendre polynomials are a set of orthogonal polynomials defined in a Hilbert space that generate a complete basis. The idea of this methodology is to represent any function of the space by using this set of basis functions. This is done by the use of inner products and the correct normalization of the Legendre polynomials. Let $f$ and $g$ be two arbitrary functions from the Hilbert space considered. Then, the inner product between these two functions is defined as:
\begin{equation}
\langle  f, g \rangle =\int_{\Omega} f({\bf x})g({\bf x}) w({\bf x})d{\bf x},
\end{equation}
where $w({\bf x})$ is a positive weighting function defined on the space domain $\Omega$. For the case of Legendre polynomials, the weighting function is a constant $w({\bf x}) = 1$, and the domain for each variable ranges between $[-1,1]$. In addition, the normalized Legendre polynomials are defined such that:
\begin{equation}
\langle  L_i, L_j \rangle =\int_{\Omega} L_i({\bf x})L_j({\bf x}) w({\bf x})d{\bf x} = \delta_{ij},
\end{equation}
where $L_i$ and $L_j$ with $\{i,j\}\in\{1,\dots,m\}$ are two given normalized Legendre polynomials from the set of basis functions selected, and $\delta_{ij}$ is Kronecker's delta.

By using this set of orthogonal multivariate polynomials, the function $u({\bf x},t)$ and the KO eigenfunctions, $\phi_i({\bf x})$, can be represented as a series expansion in terms of this set of basis functions:
\begin{subequations}\label{series}
\begin{align}
u({\bf x},t)=\sum_{\ell=1}^{\infty}c_{\ell}(t)L_{\ell}({\bf x})\approx \sum_{\ell=1}^{m}c_{\ell}L_{\ell}({\bf x})={\bf c}^T(t){\bf L}({\bf x}) = {\bf L}^T({\bf x}){\bf c}(t),\\
\phi_i({\bf x},t)=\sum_{\ell=1}^{\infty}p_{i\ell}(t)L_{\ell}({\bf x})\approx \sum_{\ell=1}^{m}p_{i\ell}L_{\ell}({\bf x})={{\bf p}_i}^T(t){\bf L}({\bf x}) = {\bf L}^T({\bf x}){\bf p}_i(t),
\end{align}
\end{subequations}
where $c_{\ell}(t)$ describes the time evolution of the function $u({\bf x},t)$ over the basis $L_{\ell}({\bf x})$ and  $p_{i\ell}(t)$ are the coefficients associated with the eigenfunction $\phi_i$ and the basis $L_{\ell}$. Moreover, ${\bf c}(t)$, ${\bf p}_{i}(t)$, and ${\bf L}$ are three column vectors containing the set of coefficients $c_{\ell}(t)$, $p_{i\ell}(t)$ and the whole set of basis functions, respectively. Note that although the series is infinite, a truncation was performed using $m$ different basis functions, and thus, this represents an approximation of the eigenfunctions.

The PDE provided in Eq.~\eqref{koopman_PDE} for the computation of the Koopman eigenfunctions can be approximated using the Galerkin method and the series expansion from Eq.~\eqref{series}. The general concept of Galerkin methods is to project the operator into the subspace $\mathcal{F}_D$ using a weighted residual technique such that the residual of Eq.~\eqref{koopman_PDE} is orthogonal to the span of $\mathcal{F}_D$. 

We start by approximating the solutions of Eq.~\eqref{koopman_PDE1} by
letting $u({\bf x},t) = {\bf c}^T(t){\bf L}({\bf x})$ be the truncated series defined by Eq.~\eqref{series}. We define the residual error of the Koopman PDE in q.~\eqref{koopman_PDE1} as:
\begin{equation}\label{residual0}
    e({\bf x},t) = \frac{d u ({\bf x},t)}{dt}-\mathcal{K}\left(u({\bf x},t)\right)
\end{equation}
where $\frac{d u ({\bf x},t)}{dt}$ is approximated by $\dot{u}({\bf x},t) \approx \dot{\bf c}^T(t){\bf L}$, with $\dot{\bf c}(t)$ a 
%constant 
vector of size $m$, that is, we want the derivative to be expressed in the same set of basis functions as the eigenfunctions. %For simplify the time dependency of the coefficient $c_{\ell}(t)$ is omitted. 
Moreover, the solution sought should be orthogonal to $\mathcal{F}_D$, being the orthogonality condition defined as:
\begin{equation}\label{residual}
\langle L_j({\bf x}),  e({\bf x},t)  \rangle =0, \quad \forall \ j \in \{1,2,\dots,m\}.
\end{equation}
The Koopman operator applied to $u({\bf x},t)$ is given by:
\begin{equation}\label{residual2}
\mathcal{K}\left(u({\bf x},t)\right) = \left(\nabla_{\bf x} {\bf L}^T{\bf c}(t)\right)^T {\bf f} =  \left({\bf f}^T\nabla_{\bf x}{\bf L}^T\right) {\bf c}(t)= {\bf f}^T {\bf L_ x^T}{\bf c}(t),
\end{equation}
where the term ${\bf L_ x^T}=\nabla_{\bf x}{\bf L}^T$ is defined as: 
\begin{equation}\label{derivatives}
{\bf L_ x^T}=\left(\begin{array}{ccc} 
    \displaystyle\frac{\partial }{\partial x_1}L_1({\bf x}) & \cdots & \displaystyle\frac{\partial }{\partial x_d}L_1({\bf x}) \\ 
    \vdots & \ddots & \vdots\\
    \displaystyle\frac{\partial }{\partial x_1}L_m({\bf x}) & \cdots& \displaystyle\frac{\partial }{\partial x_d}L_m({\bf x})\end{array}\right).
\end{equation}
It is important to note that all of the terms in the equation above can be calculated in closed-form since both ${\bf L}$ and ${\bf L_ x^T}$ are only comprised by polynomials. Therefore, it is possible to apply Eq.~\eqref{residual2} into Eq.~\eqref{residual0} to obtain:
\begin{eqnarray}
    \langle  L_j({\bf x}), e({\bf x},t)  \rangle & = & \langle  L_j({\bf x}), {\bf L}^T({\bf x})\dot{\bf c}(t) \rangle - \langle  L_j({\bf x}),{\bf f}^T {\bf L_ x^T}{\bf c}(t) \rangle = \nonumber \\
    & = & \langle  L_j({\bf x}), {\bf L}^T({\bf x}) \rangle \dot{\bf c} - \langle  L_j({\bf x}),{\bf f}^T {\bf L_ x^T} \rangle {\bf c}(t)  = 0,
\end{eqnarray}
Above we define a set of $m$ equations for each $L_j({\bf x})$, we now gather these equations into a matrix vector representation to obtain:
\begin{equation} \label{time_matrix_form}
    \frac{d {\bf c}(t)}{dt} =G^{-1}K{\bf c}(t),
\end{equation}
where:
\begin{equation} 
    G_{ij}= \langle  L_i({\bf x}),   L_j({\bf x})\rangle,
\end{equation}
However, due to the orthogonality of the basis functions we have that $G=I_{m\times m}$. Therefore, the $K$ matrix is the matrix representation of the KO over the basis $L_i({\bf x})$, a $m \times m$ sized matrix whose components are:
\begin{equation} \label{kmatrix}
    K_{ij} = \langle L_i({\bf x}), {\bf f}^T \nabla_{\bf x}L_j({\bf x})\rangle.
\end{equation}
Equation \eqref{time_matrix_form}  determines the time evolution of $u({\bf x},t)$ by solving for ${\bf c}(t)$. It is a set of coupled ordinary differential equations that can be solved using the matrix exponential, ${\bf c}(t)=\exp \left(Kt\right){\bf c}(t_0)$ where ${\bf c}(t_0)$ is the initial coefficient vector. 
%\begin{equation}
 %   \langle {\bf p}_i{\bf L}^T({\bf x}), L_j({\bf x}) \rangle = {\bf c_i} \langle {\bf L_ x^T}{\bf f}({\bf x}), L_j({\bf x}) \rangle = p_{ij},
%\end{equation}
%and:
%\begin{equation}
%    \langle  \lambda_i\sum_{\ell=1}^{\infty}c_{i\ell}L_{\ell}({\bf x}), L_j({\bf x}) \rangle = \lambda_ic_{ij},
%\end{equation}
%due to the orthogonality of the basis functions. This means that we have to impose that:
%\begin{equation}
   % p_{ij} = {\bf c_i} \langle {\bf L_ x^T}{\bf f}({\bf x}), L_j({\bf x}) \rangle = \lambda_{ij}c_{ij},
%\end{equation}
%Alternatively, we could have derived the time evolution of the basis functions $L_i({\bf x})$ through the following expression 
%%\frac{d L_i({\bf x})}{dt}={ f}_1\left({\bf x}\right)\frac{\partial }{\partial x_1} L_i({\bf x}) +\cdots+  { f}_d\left({\bf x}\right)\frac{\partial }{\partial x_d}L_i({\bf x}),
%\end{equation}
%
%Using the Galerkin method we can 

Note that by using Eq.~\eqref{time_matrix_form} we can obtain the solution for any state of the system. In order to do that, we start by defining the initial set of coefficient vectors as $C_0=[{\bf c}_0^1,\cdots,{\bf c}_0^m]$, where $C_0\in\mathbb{R}^{m \times d}$, ${\bf c}_0^i\in\mathbb{R}^{m \times 1}$, and ${\bf c}_0^i={\bf c}_i(t_0)$ is the initial coefficient vector for the $i^\text{th}$ coordinate, $x_i$, and $x_i(t_0)=L^T({\bf x}){\bf c}_0^i$. Then, the initial condition can be expanded in $L_i$ as ${\bf x}^T(t_0)=L^T({\bf x})C_0$ and the solution for each of these coefficient vectors is given by $C(t)=\exp(Kt)C_0$. These results can be merged to obtain the approximate evolution of the state:
\begin{equation} \label{decomposition}
    {\bf x}(t)=C_0^T\exp(K^Tt)L({\bf x}_0),
\end{equation}
where the transpose of $C_0$ and $K$ is needed since ${\bf x}(t)$ is a column vector. This solution for the state will be expanded on in the next section to use the KO eigenfunctions. Note that the coefficient $C_0$ can be computed analytically using the Legendre polynomials basis functions. Additionally, $K$ is also computed analytically, and $L({\bf x}_0)$ are the collection of normalized Legendre basis functions up to a given maximum order. Therefore, the solution in Eq.~\eqref{decomposition} is an approximate analytical solution to Eq.~\eqref{problem} given the initial condition ${\bf x}_0$ at time $t_0$.

If we repeat this process for Eq.~\eqref{koopman_PDE} and consider all the basis functions with $j\in\{1,\dots,m\}$, we can show that the eigenvectors of the matrix $K$ are the coefficient vectors of the KO eigenfunctions with the same corresponding eigenvalue. This relationship is given by
\begin{equation} \label{decomposition2}
    P K =  \Lambda P,
\end{equation}
where $P$ is the matrix containing all the eigenvectors of the matrix $K$, that is, these eigenvectors are the coefficients $p_{ij}$ of the KO eigenfunctions. Therefore, the KO eigenfunctions are then given by 
\begin{equation} \label{decomposition3}
    \Phi({\bf x}) =  L^T({\bf x})P
\end{equation}
Additionally, $\Lambda$ is a diagonal matrix of eigenvalues $\lambda_i$ of the matrix $K$, where the eigenvalue $\lambda_i$ relates to the $i$th eigenfunction. Note that Eq.~\eqref{decomposition} is in fact defining an eigendecomposition of matrix $K$, where $P$ is the matrix containing the left eigenvectors of $K$, and $\Lambda$ the associated eigenvalues. Therefore, it is possible to first obtain $K$ in closed-form solution from Eq.~\eqref{kmatrix}, and then perform the eigendecomposition to obtain $P$, and thus, the approximated eigenfunctions $\phi_i$ of the system through Eq.~\eqref{series}.

\subsection{Koopman Operator Theory: Linear System}

Let us consider the Koopman eigenvalues and eigenfunctions of a Linear Time Invariant (LTI) system. LTI systems are a particular case of the initial value problem where it is possible to compute the Koopman eigenfunctions analytically; thus, these systems are referred to as solvable. Consider the following LTI system:
 \begin{equation}\label{linear}
\dot{\bf x}=\Xi{\bf x},
\end{equation}
with ${\bf x}\in\mathbb{R}^d$ and $\Xi\in\mathbb{R}^{d\times d}$. The eigendecomposition of this linear system yields complete information about the underlying dynamics provided that $\Xi$ has a complete set of eigenvectors, left and right eigenvectors ${\bf w}_i$ and ${\bf v}_i$ respectively, with corresponding eigenvalues $\zeta_i$. Given that the dynamics are linear, it is expected that the Koopman approach should contain the eigendecomposition of the matrix $\Xi$.
To show this, the KO for the linear system is
$
\frac{d}{dt}\phi({\bf x})=\mathcal{K}\left(\phi({\bf x})\right)=(\Xi{\bf x})^T\left[ \nabla\phi({\bf x})\right]
$. Assuming that $\Xi$ has a complete set of eigenvectors, there are $d$ left eigenvectors, ${\bf w}_i$, that satisfy ${\bf w}^T_i\Xi = \mu_i{\bf w}^T_i$. It is easy to verify that $\phi_i({\bf x})={\bf w}_i^T{\bf x}$ is an eigenfunction of the KO for the linear system in Eq.~\eqref{linear} by computing the KO
\begin{equation}
\mathcal{K}\left(\phi_i({\bf x})\right)={\bf x}^T \Xi ^T\nabla\left({\bf w}^T_i{\bf x}\right)={\bf x}^T \Xi ^T{\bf w}=\zeta_i\phi_i({\bf x}).
\end{equation}
Therefore, each one of the eigenvectors of the matrix $\Xi$ can be used to construct a first order polynomial in the state variable, $\phi_i({\bf x})={\bf w}_i^T{\bf x}$, which is a KO eigenfunction for this linear system. Then, thanks to the algebraic property of the KO, an infinite number of eigenfunctions for this linear system can be expressed as a polynomial function of the state (from the product of the first order polynomials $\phi_i({\bf x})={\bf w}_i^T{\bf x})$.
Thus, the function
\begin{equation}
\phi_\ell({\bf x})=\prod_{i=1}^{d}\left({\bf w}^T_i{\bf x}\right)^{n_{i,\ell}}
\end{equation}
is an eigenfunction of the KO with associated eigenvalue, $\lambda_{\ell}=\sum_{i}^{d}n_{i,\ell}\zeta_i$.

%*****************************************************************************
%*****************************************************************************
%*****************************************************************************

\subsection{Koopman modes}

The Koopman modes are the representation of the observables into the set of eigenfunctions ${\bf \Phi}$ of the system. Let ${\bf g}({\bf x})$ be the set of observables in which we are interested. These can be any function of the original variables ${\bf x}$, including the states themselves, called the identity observables. These observables can be expressed as a series expansion in the set of eigenfunctions obtained before:
\begin{equation}
    g_i({\bf x}) \approx \sum_{\ell=1}^m b_{i\ell}\phi_{\ell}({\bf x}),
\end{equation}
where $b_{i\ell}$ is a set of constant coefficients that can be computed by projecting the observable into the set of basis functions:
\begin{equation}
    b_{i\ell} = \left\langle g_i, \phi_{\ell} \right\rangle.
\end{equation}
This means that ${\bf g}({\bf x})$ can also be represented in matrix notation as:
\begin{equation}\label{observable}
    {\bf g}({\bf x}) = B {\bf \Phi}({\bf x}),
\end{equation}
where $B$ is a matrix of size $q \times m$ containing the coefficients $b_{i\ell}$. The parameter $q$ represents the number of observables considered, for instance, in the particular case of the identity observables, $q = d$, that is, the number of dimensions of the original system of equations.  Therefore, it is possible to obtain the evolution of any observable ${\bf g}({\bf x})$ by means of the evolution of the eigenfunctions of the system. Particularly, from Eq.~\eqref{eigen_time}:
\begin{equation}
    {\bf g}({\bf x}(t)) = B {\bf \Phi}({\bf x}(t)) = B \exp(\Lambda t){\bf \Phi}({\bf x}(t_0)),
\end{equation}
or expressed in terms of the basis functions used:
\begin{equation} \label{solution}
    {\bf g}({\bf x}(t)) = B {\bf \Phi}({\bf x}(t)) = B \exp(\Lambda t) P {\bf L}({\bf x}(t_0)),
\end{equation}
where ${\bf x}(t_0)$ are the values of the states in the initial condition. 

Note that the observables can also be represented by the set of basis functions ${\bf L}$:
\begin{equation} \label{obs}
    {\bf g}({\bf x}) = A {\bf L}({\bf x}) \rightarrow g_i({\bf x}) \approx \sum_{\ell=1}^m a_{i\ell}L_{\ell}({\bf x}), \quad \text{where} \quad a_{i\ell} = \left\langle g_i, L_{\ell} \right\rangle,
\end{equation}
where $A_{ij}=a_{ij}$, whose solution under the system of differential equations is:
\begin{equation}\label{KO_solution}
    {\bf g}({\bf x}(t)) = A {\bf L}({\bf x}(t)) = A P^{-1} \exp(\Lambda t) P {\bf L}({\bf x}(t_0)).
\end{equation}
Additionally, it is important to note that the result from Eq.~\eqref{solution} can also be used to obtain the state transition matrix of the system. In that regard, it is important to note that if the identity observable is used, that is, ${\bf g}({\bf x}) = {\bf x}$ then we have that $C_0^T=A$, and thus:
\begin{equation} \label{state_trans}
\displaystyle\frac{\partial {\bf x}(t_f)}{\partial {\bf x}(t_0)} = B\exp(\Lambda t_f)P\frac{\partial {\bf L}({\bf x}(t_0))}{\partial {\bf x}(t_0)}.
\end{equation}
Higher-order state transition matrix terms can also be computed from Eq.~\eqref{solution}. In this work, we make use of Eq.~\eqref{state_trans} to approximate periodic solutions to the RTBP using a simple differential corrector method.

%*****************************************************************************
%*****************************************************************************
%*****************************************************************************

\section{Dynamics for the Motion near $\mathcal L_1$ or $\mathcal L_2$ in the Circular Restricted Three Body Problem}

The circular restricted three body problem (RTBP) equations can be expressed using the following form:
\begin{subequations}
\begin{align}
\ddot{x}&=\frac{\partial \Omega}{\partial x}+2\dot{y},\\
\ddot{y}&=\frac{\partial\Omega }{\partial y}-2\dot{x},\\
\ddot{z}&=\frac{\partial \Omega }{\partial z},
\end{align}
\end{subequations}
where $\Omega=1/2\left(x^2+y^2\right)+(1-\mu)/r_1+\mu/r_2$, $\mu$ is the mass of the secondary divided by the mass of the primary object, and $r_1$ and $r_2$ are the distances to the primary and secondary masses, respectively. The distance from the libration points, $\mathcal L_j$ to the primary body, $\gamma$, is found by solving the Euler quintic equation:
\begin{subequations} \label{eqgamma}
\begin{align}
\gamma_j^5&\mp \left(3-\mu\right)\gamma_j^4+\left(3-2\mu\right)\gamma_j^3-\mu\gamma_j^2\pm2\mu\gamma_j-\mu=0, &\quad & \text{if} \quad j=1,2;\\
\gamma_j^5&+ \left(2-\mu\right)\gamma_j^4+\left(1-2\mu\right)\gamma_j^3-\left(1-\mu\right)\gamma_j^2-2\left(1-\mu\right)\gamma_j-\mu-1=0, &\quad & \text{if} \quad j=3.
\end{align}
\end{subequations}

The circular Restricted Three-Body Problem (RTBP) model that used for this work follows the work from Richardson~\cite{richardson1980analytic}. In particular, the equations of motion for a satellite moving near the $\mathcal L_1$ or $\mathcal L_2$ points can be obtained by translating the origin to the location of $\mathcal L_1$ or $\mathcal L_2$. This coordinate transformation is given by the following expressions:
\begin{subequations} \label{coord}
\begin{align} 
\bar{x}&=\frac{x-1+\mu\pm\gamma}{\gamma},\\
\bar{y}&=\frac{y}{\gamma},\\
\bar{z}&=\frac{z}{\gamma},
\end{align}
\end{subequations}
where $\mu$ is the mass ratio of the primary to the secondary object, and the $\pm$ symbol relates to the dynamic of the system in the following manner: the upper sign is a dynamic about $\mathcal L_1$, while the lower sign is for $\mathcal L_2$.
Additionally, the variables $x$, $y$, and $z$ are the position coordinates for the RTBP, with the origin at the primary body, while $\bar{x}$, $\bar{y}$, and $\bar{z}$ are the position variables in the shifted and scaled coordinates. In $\bar{x}$, $\bar{y}$, and $\bar{z}$, the new origin is either the $\mathcal L_1$ or the $\mathcal L_2$ point, provided that these variables are normalized such that the distance between $\mathcal L_1$ or $\mathcal L_2$ and the primary is 1. This scaled coordinate system was introduced by Richardson~\cite{richardson1980analytic} and was shown to improve the numerical properties of the RTBP model. For convenience of notation, $x$, $y$, $z$ will refer to $\bar{x}$, $\bar{y}$, and $\bar{z}$ variables henceforth. 

The full nonlinear RTBP equations can be expanded in terms of polynomials as Richardson~\cite{richardson1980analytic} showed. Additionally, Ref.~\cite{richardson1980analytic} showed that using Legendre polynomials, $P_n$, for expanding the RTBP equations can lead to computational advantages. The main nonlinear terms in the RTBP model involve $\frac{1-\mu}{r_1}+\frac{\mu}{r_2}$, where $r_1$ and $r_2$ are the distances to the primary and secondary objects, can be approximated using the following relation:
\begin{equation}
\frac{1}{\sqrt{(x-A)^2+(y-B)^2+(z-C)^2}}=\frac{1}{D}\sum_{n=0}^{\infty}\left(\frac{\rho}{D}\right)^nP_n\left(\frac{Ax+By+Cz}{D\rho}\right),
\end{equation}
where $D^2=A^2+B^2+C^2$ and $\rho^2=x^2+y^2+z^2$. Therefore, following the development in Ref.~\cite{richardson1980analytic}, the nonlinear RTBP equations can be expressed with respect to the $\mathcal L_1$ or $\mathcal L_2$ points as:
\begin{subequations}\label{RTBP}
\begin{align}
\ddot{x}-2\dot{y}-(1+2c_2)x&=\frac{\partial}{\partial x}\sum_{n\geqslant 3} c_n \rho^n P_n\left(\frac{x}{\rho}\right),\\
\ddot{y}+2\dot{x}+(c_2-1)y&=\frac{\partial}{\partial y}\sum_{n\geqslant 3} c_n \rho^n P_n\left(\frac{x}{\rho}\right),\\
\ddot{z}+c_2z&=\frac{\partial}{\partial z}\sum_{n\geqslant 3} c_n \rho^n P_n\left(\frac{x}{\rho}\right),
\end{align}
\end{subequations}
Note that the left-hand side contains only linear terms. The coefficients $c_n$ are given by:
\begin{equation}\label{c_equation}
c_n=\frac{1}{\gamma^3}\left((\pm)^n\mu+(-1)^n\frac{(1-\mu)\gamma^{n+1}}{(1\mp\gamma)^{n+1}}\right),
\end{equation}
where the upper sign is for $\mathcal L_1$ and the lower one for $\mathcal L_2$. This work considers the Sun-Earth system. In the case of $\mathcal L_1$, the values of the problem constants are $\gamma=0.0099703255$ and
$\mu=3.0034106426\times 10^{-6}$, which can then be used with Eq.~\eqref{c_equation} to compute $c_2 = 4.060821911$, $c_3 = 3.019929488$, and $c_4 = 3.030412038$.

This model can be calculated using an elegant recursive formula~\cite{koon2000dynamical}. In particular, Legendre polynomials can be calculated recursively using the well-known three-term recursive formula, which can be in turn used to define a recursion for the RTBP model. Consider the following function: 
\begin{equation}
T_n(x,y,z) =\rho^n P_n\left(\frac{x}{\rho}\right),
\end{equation}
where $T_n$ is a homogeneous polynomial of degree $n$ in variables $x$, $y$, and $z$. Using the Legendre polynomial three-term recursive formula, it can be shown that:
\begin{equation}
T_n =\frac{2n-1}{n}xT_{n-1}-\frac{n-1}{n}\left(x^2+y^2+z^2\right)T_{n-2},
\end{equation}
where the recursion starts with $T_0 = 1$ and $T_1 = x$. 

Legendre polynomials had already been applied to approximate the inverse potential for the N-body problem~\cite{aarseth1967collisionless}, with particular attention to the circular RTBP, leading to multiple Legendre polynomial constructions (listed in~\cite{laskar2010explicit}) . Richardson analyzed the presented polynomial approach and best suited the representation for the RTBP, truncating Eq.~\eqref{RTBP} at order $n=3$. This work uses this formula to define the polynomial model of the RTBP, and it is henceforth referred to as the Richardson's formulation. The Richardson's model used for the comparisons in the numerical applications presented in this paper involves truncating Eq.~\eqref{RTBP} at order $n=10$.

\subsection{Hamiltonian in Normal Form}

Jorba and Masdemont~\cite{jorba1999dynamics} provided a summary of the normal form computation for the RTBP model and this section follows the notation and process given in that work. The equations of motion of the CRTBP following this model are:
\begin{equation}\label{H1}
H=\frac{1}{2}\left(p_x^2+p_y^2+p_z^2\right)+yp_x-xp_y-\sum_{n\geq2} c_n \rho^n P_n\left(\frac{x}{\rho}\right),
\end{equation}
where the variables are defined as pseudo momenta, $p_x=\dot{x}-y$, $p_y=\dot{y}+x$, and $p_z=\dot{z}$. Then, the normal form for this system is found by using a symplectic linear change of variables that utilizes the eigenvectors of the linearized matrix to simply Eq.~\eqref{H1}. The linearized system is given by 
\begin{equation}
\dot{\bf v}=M{\bf v},
\end{equation}
where ${\bf v}=[{x}, {y}, {z}, p_x, p_y, p_z]^T$ and $M$ is determined by evaluating the sum in Eq.~\eqref{RTBP} for only the $n=2$ terms, leading to:
\begin{equation} \label{M}
M=\begin{bmatrix}
0 & 1 & 0 & 1 & 0 & 0\\
-1 & 0 & 0 & 0 & 1 & 0\\
0 & 0 & 0 & 0 & 0 & 1\\
2c_2 & 0 & 0 & 0 & 1 & 0\\
0 & -c_2 & 0 & -1 & 0 & 0\\
0 & 0 & 0 & 0 & 0 & -c_2\\
\end{bmatrix}.
\end{equation}
Note that the $z$ dimension is decoupled from $x$ and $y$. Therefore, the characteristic polynomial for the $x$-$y$ motion is given by:
\begin{equation}
p(\lambda)=\lambda^4+\left(2-c_2\right)\lambda^2+\left(1+c_2-2c_2^2\right)=0,
\end{equation}
where the solutions for the characteristic polynomial in the $x$-$y$ direction are:
\begin{equation}
\omega_1^2=\frac{c_2-2-\sqrt{9c_2^2-8c_2}}{2}, \quad \lambda_1^2=\frac{c_2-2+\sqrt{9c_2^2-8c_2}}{2}.
\end{equation}
Under the assumption that $\mu>0$, the $x$-$y$ motion characteristic equation has four roots, two real and two imaginary given by $(\pm \lambda_1, \pm  \omega_1)$. The $z$ direction motion is characterized by two imaginary eigenvalues $\pm \omega_2$ given by $\omega_2^2=c_2$. The linear behavior near the collinear libration point is that of a saddle×center×center. The eigenvectors for the $x$-$y$ coupled linearized system are given by ${\bf e}_{\lambda}=[2\lambda,\lambda^2-2c_2-1,0,\lambda^2+2c_2+1,\lambda^3\left(1-2c_2\right),0]^T$. The $z$ eigenvector is given by ${\bf e}_{z}=[0,0,1,0,0,0]^T$ and ${\bf e}_{\dot{z}}=[0,0,0,0,0,1]^T$. By using the eigenvector, we can form a symplectic transformation $C$ that normalizes the $x$-$y$ motion such that $C^TJC=J$, where $J$ is given by 
\begin{equation}
J=\begin{bmatrix}
0_{3\times3} & I_{3\times3}\\
-I_{3\times3} & 0_{3\times3} 
\end{bmatrix}.
\end{equation}
Additionally, using the normalization discussed in Jorba and Masdemont~\cite{jorba1999dynamics}, the symplectic transformation can be expressed as:
\begin{equation} \label{C}
C=\begin{bmatrix}
\frac{2\lambda_1}{s_1} & 0 & 0 & -\frac{2\lambda_1}{s_1} & \frac{2\omega_1}{s_2} & 0\\
\frac{\lambda_1^2-2c_2-1}{s_1} & \frac{-\omega_1^2-2c_2-1}{s_2} & 0 & \frac{\lambda_1^2-2c_2-1}{s_1} & 0 & 0\\
0 & 0 & \frac{1}{\sqrt{\omega_2}} & 0 & 0 & 0\\
\frac{\lambda_1^2+2c_2+1}{s_1} & \frac{-\omega_1^2+2c_2+1}{s_2} & 0 & \frac{\lambda_1^2+2c_2+1}{s_1} & 0 & 0\\
\frac{\lambda_1^3+\left(1-2c_2\right)\lambda_1}{s_1} & 0 & 0 & \frac{-\lambda_1^3-\left(1-2c_2\right)\lambda_1}{s_1} & \frac{-\omega_1^3+\left(1-2c_2\right)\omega_1}{s_2} & 0\\
0 & 0 & 0 & 0 & 0 & \sqrt{\omega_2}\\
\end{bmatrix}
\end{equation}
where $s_1=\sqrt{2\lambda_1\left(\left(4+3c_2\right)\lambda_1^2+4+5c_2-6c_2^2\right)}$ and $s_2=\sqrt{\omega_1\left(\left(4+3c_2\right)\omega_1^2-4-5c_2-6c_2^2\right)}$. Applying a transformation such that ${\bf v}'=C{\bf v}$ is the new set of variables, we can rewrite the Hamiltonian in the following form: 
\begin{equation}\label{H2}
H_2=\lambda_1xp_x+\frac{\omega_1}{2}\left(y^2+p_y^2\right)+\frac{\omega_2}{2}\left(z^2+p_z^2\right)-\sum_{n\geq2} c_n \rho^n P_n\left(\frac{x}{\rho}\right),
\end{equation} % y^2+py^2 should be z^2+pz^2 (copy-paste casualty), good catch!
where $x$ and $\rho$ can be computed in terms of the new variable using the transformation $C$.
Finally, the complex normal transformation can be achieved with the following transformation:
\begin{subequations}\label{CRTBP}
\begin{align}
x=q_1, \quad y=\frac{q_2+\sqrt{-1}p_2}{\sqrt{2}}, \quad z=\frac{q_3+\sqrt{-1}p_3}{\sqrt{2}}\\
p_x=p_1, \quad p_y=\frac{\sqrt{-1}q_2+p_2}{\sqrt{2}}, \quad p_z=\frac{\sqrt{-1}q_3+p_3}{\sqrt{2}}
\end{align}
\end{subequations}%According to Jorba and Masdemont, py term is incorrect.good catch!
This final transformation can be written in matrix form and we will denote the complexification transformation as $C_2$. Therefore the final variable in the complex normal form is given by ${\bf v}''=C_2C{\bf v}$. The complex normal form Hamiltonian is then given by:
\begin{equation}\label{H3}
H_3=\lambda_1 q_1 p_1 +\sqrt{-1}\omega_1 q_2 p_2 +\sqrt{-1} \omega_2 q_3p_3-\sum_{n\geq2} c_n \rho^n P_n\left(\frac{q_1}{\rho}\right),
\end{equation}
where $q_1$ and $\rho$ can be computed in term of the new variable using the transformation $C_2C$. Equation \eqref{H3} can then be used with Hamilton's equation, $\dot{\bf v}''=J\nabla H_3$, to compute the equations of motion in the ${\bf v}''$ variables. The Hamiltonian $H_3$ is used to compute the KO of the RTBP since it decouples the variables in the linear leading terms, and it also decouples the effects of the stable, unstable, and periodic manifolds. Additionally, we introduce an additional scaling for the variables, where each variable is multiplied by a small constant, $\alpha$. The scaling by $\alpha$ further reduces the magnitude of the coefficients $c_n$ for the higher-order terms in the series shown in Eq.~\eqref{H3}, increasing the accuracy of the Koopman approximation, and minimizing numerical problems. A value of $\alpha = 0.01$ has been chosen for the results presented in the following sections.

\subsection{Complex Computation of Koopman Operator}
The state vector in the original reference frame, i.e. ${\bf x}(t)$, has been translated and scaled to be consistent with Richardson's Hamiltonian formulation, $H$, which depends of a different state vector ${\bf v}(t)$. Afterwards, two separate transformations rewrite the Hamiltonian into its normal form, $H_3$, where the state of the transformed system is ${\bf v}(t)'' =[{q_1}, {q_2}, {q_3}, p_1, p_2, p_3]^T = C_2 C{\bf v}(t) $, which can take complex outcomes. The final Hamiltonian formulation has complex components and therefore, the Koopman operator will as well. We compute the Koopman operator by dividing the resulting dynamical system into complex and real parts given by
\begin{equation}
{\bf f}({\bf v''})={\bf f}_\text{Re}({\bf v''})+\sqrt{-1}{\bf f}_\text{Img}({\bf v''}),
\end{equation}
where ${\bf f}_\text{Re}({\bf v''})$ and ${\bf f}_\text{Img}({\bf v''})$ are the real and imaginary parts of the dynamic function, respectively. The dynamics ${\bf f}({\bf v''})$ are evaluated by applying Hamilton's equation to the complex Hamiltonian $H_3({\bf v''})$, Eq.~\eqref{H3}. Then the Koopman operator complex computation can be performed with Legendre polynomials using the follow expression:
\begin{equation} \label{kmatrix2}
    K_{ij} = \langle L_i({\bf v''}), {\bf f}_\text{Re}^T \nabla_{\bf v''}L_j({\bf v''})\rangle+\sqrt{-1}\langle L_i({\bf v''}), {\bf f}_\text{Img}^T \nabla_{\bf v''}L_j({\bf v''})\rangle.
\end{equation}
The polynomial approximation of the dynamics leads to a clean separation between real and imaginary monomials in the equations of motion. As such, each entry of the Koopman matrix is evaluated by adding the integrals (through the inner products) of each single monomial of $\bf f (\bf v'')$ multiplied by the Legendre polynomials. There is no need to separate the Legendre polynomials into their real and imaginary parts since it is assumed that the state variable $\bf v''$ is complex. Consequently, the basis functions cover all possible outcomes in the state space, and provide a complex output given a complex input. The division of the dynamics into ${\bf f}_\text{Re}({\bf v''})$ and ${\bf f}_\text{Img}({\bf v''})$ simplifies the evaluation of the Koopman matrix by working individually on each single monomial of the polynomial representation of the dynamics and then summing all the contributions together. 

The influence of a complex representation of the ODEs affects, also, the evaluation of the observable matrix $A$ from Eq.~\eqref{obs}. Each entry of $A$ is calculated as the inner product between the observable function $\bf g(\bf v'')$ and the basis functions $\bf L(\bf v'')$. In order to obtain the time evolution of the state of the system, we are required to select the identity observable. However, due to the multiple transformations applied to the Hamiltonian, the identity observable must be transformed to be consistent with the new variable set ${\bf v}''= C_2 C {\bf v}$, which identifies the ``normal form" representation of the state. Therefore, the observable matrix A is evaluated by separating real and complex contributions as for the Koopman matrix. 
\begin{equation}
    a_{ij} = \left\langle g_{i,\text{Re}}({C_2 C\bf v}), L_{j} ({\bf v''})\right\rangle + \sqrt{-1}\langle  g_{i,\text{Img}}({C_2 C\bf v}), L_j({\bf v''})\rangle.
\end{equation}

Lastly, given the initial condition of the state ${\bf x}(t_0)$ in the original reference frame, the state vector is translated and scaled according to Eq.~\eqref{coord}, giving ${\bf v}_0$. Afterwards, the initial condition undergoes the inverse of transformation $C_2 C$ before being evaluated by the basis functions during the calculation of the Koopman solution. Thus, the Legendre polynomials complex evaluation is applied only to get the final solution of the system, and not during the integration in the inner product. In this way, Eq.~\eqref{KO_solution} outputs directly the state into its not-transformed representation $({\bf v})$, as it was calculated directly from the Hamiltonian defined in Eq.~\eqref{H1}, working with observable ${\bf g}({\bf v})$ over ${\bf g}({\bf v''})$. This property can be appreciated by analyzing the matching of the initial condition with the Koopman solution. Given the identity observable, at time $t_0$ the solution is:
\begin{align}
    {\bf g}({\bf v}_0) &= A P^{-1} \exp(\Lambda t_0) P {\bf L}((C_2 C)^{-1} {\bf v}_0)  \\
    &= A P^{-1} P {\bf L}((C_2 C)^{-1} {\bf v}_0) \\
    &= A {\bf L}((C_2 C)^{-1} {\bf v}_0);
\end{align}
where, for the $i$th observable, calling with $A_i$ the $i$th row of matrix $A$:
\begin{align}
     g_i({\bf v}_0) &=  A_i {\bf L}((C_2 C)^{-1} {\bf v}_0)\\
     &= \sum_{j=1}^{m} \bigg(  \left\langle g_{i,\text{Re}}({C_2 C\bf v}), L_{j} ({\bf v''})\right\rangle + \sqrt{-1}\langle  g_{i,\text{Img}}({C_2 C\bf v}), L_j({\bf v''})\rangle  \bigg)  L_j((C_2 C)^{-1} {\bf v}_0);
\end{align}
where, once again, $m$ is the number of basis functions. The transformations in the observable, $C_2 C$, and in the basis functions, $(C_2 C)^{-1}$, simplify themselves, giving, as solution, the state of the system as if it were evaluated from the Hamiltonian $H$. The evaluation of the Legendre polynomials ${\bf L}((C_2 C)^{-1} {\bf v}_0)$ is a simple variable substitution, where the transformed state given as input is complex and it leads to a complex output of the polynomial evaluation.  

After applying KO to the Hamiltonian in normal form and obtaining the final solution as if from $H$ (thanks to some mathematical manipulations), the state of the system is scaled and translated back to the original reference frame by inverting the coordinate transformation Eq.~\eqref{coord}, getting ${\bf x}(t)$ from ${\bf v}(t)$. As the final step, the velocity is derived directly from the pseudo momenta: $\dot{x} = p_x + y$, $\dot{y} = p_y - x$, and $\dot{z} = p_z$.

%%%%%%%%%%%%%%%%%%%%%%%%%%%%%%%%%%%%%%%%%%%%%%%%%%%%%%%%%%%%%%%%%%%%%%%%%%%%%%%%%%%%%%%%%%%%%%

\section{Lyapunov Orbit Approximation Around $\mathcal L_1$}
The Koopman approximation of the dynamics is here applied for periodic orbits (POs) around the libration points $\mathcal L_1$ and $\mathcal L_2$. In the following paragraphs, the Koopman solution is shown to be an accurate propagator for Lyapunov and Halo orbits. The Koopman Operator theory is analyzed in terms of eigenvalues and eigenfunctions: the frequencies and modes of the system are observed, and a comparison among different expansion orders is presented. Lastly, the Koopman theory is incorporated in a station-keeping application, where the Koopman solution is the contribution for a newly-developed estimator and controller.

The KO solutions proposed in these sections have been achieved using a 10th order Richardson model to represent the dynamics of the RTBP. The same polynomial model, integrated numerically (with a Runge Kutta 4/5 or 7/8 propagator), would diverge. The Richardson solution is not able to achieve an accurate approximation of either the Lyapunov or Halo orbits. The accuracy of Richardson's polynomial solution decreases as the time propagation continues and Richardson's approximation of the periodic orbits diverges before achieving a full revolution.

\begin{figure}[h!]
	\centering
	\includegraphics[width=0.93\textwidth]{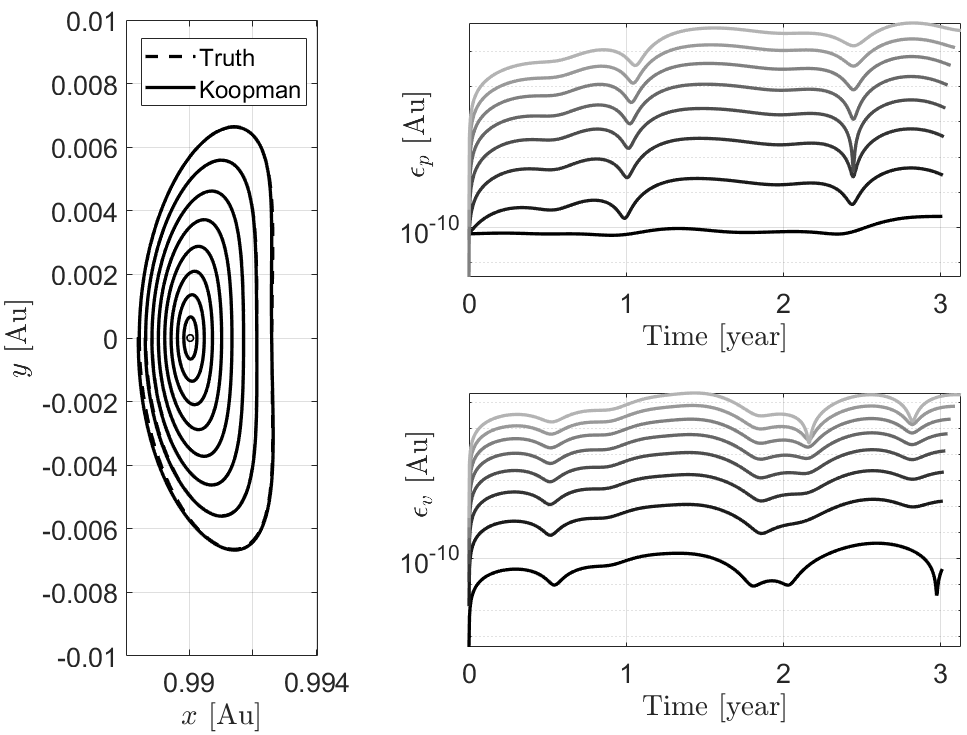}
	\caption{Koopman solution (Order 5) of the $\mathcal L_1$ Lyapunov orbits (left) and relative position (top) and velocity (bottom) error magnitudes.  }
	\label{fig:L1}
\end{figure}

The KO is first applied to solve the Lyapunov orbits around the first Lagrangian point, $\mathcal L_1$. These families of periodic orbits are planar and lie in the $(x,y)$ plane.  Figure~\ref{fig:L1} shows the Koopman approximation, represented by a set of continuous curves, of the solution of the RTBP dynamics. As can be seen, the continuous lines overlap the truth solution, which is represented by dashed lines and was evaluated numerically by a Runge-Kutta integration scheme of the original dynamics (without any approximation). 

The right side of Figure~\ref{fig:L1} reports the errors for the position and velocity,top and bottom of the figure respectively. The errors are evaluated as norms:
\begin{align}
    \epsilon_p = \sqrt{(x_{KO} - x_T)^2 + (y_{KO} - y_T)^2 + (z_{KO} - z_T)^2}; \label{eq:pos}\\
    \epsilon_v = \sqrt{(\dot x_{KO} - x_T)^2 + (y_{KO} - y_T)^2 + (z_{KO} - z_T)^2};
\end{align}
where the subscript $KO$ refers to the Koopman approximation, while the subscript $T$ is the Runge-Kutta solution.

By analyzing the POs from Figure~\ref{fig:L1}, it is noted that the accuracy of the solution decreases as the initial condition moves away from the equilibrium point. The KO approximates the orbits as a linear combination of the eigenfunctions of the system. Therefore, in order to increase accuracy for initial conditions that are farther from the equilibrium point, the order of the KO should be increased as well. The figure shows the solution of the KO using basis functions up to order 5, which corresponds to a total of 462 different eigenfunctions. However, the authors have observed~\cite{koopmanzonal} that the KO has similar performance characteristics as perturbation theory methods, where a robust solution is achieved by considering small perturbations of the original system and, therefore, high accuracy levels for far initial conditions are harder to achieve. As the figure shows, the position and velocity errors of the KO solutions increase the farther the initial condition of the PO is from the $\mathcal L_1$ equilibrium point. The gray lines in the error plots, represented with a gradient getting darker the closer the initial condition is to the $\mathcal L_1$ point, show a similar behavior along with the whole orbit revolution, and they settle on different levels of accuracy. Therefore, for small Lyapunov orbits, the KO solution has shown to be an extremely accurate approximation.

\begin{figure}[h!]
	\centering
	\includegraphics[width=1.0\textwidth]{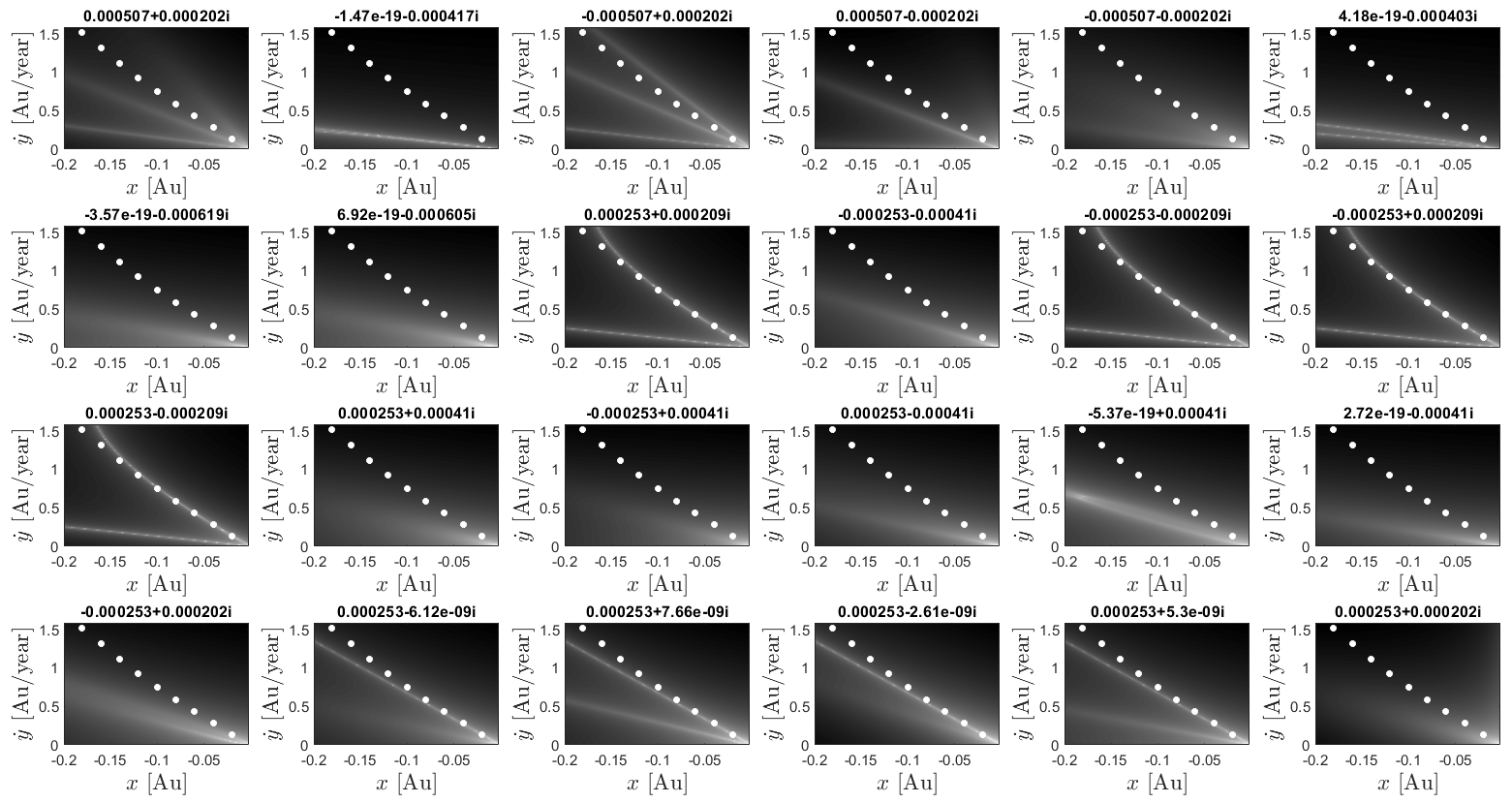}
	\caption{First 24 eigenfunctions of the 3rd order KO for the RTBP and Lyapunov initial conditions.   }
	\label{fig:eig}
\end{figure}

The eigenfunction analysis and the study of the minimum points (minima) are important when seeking stability and periodic orbits in a chaotic application, such as the RTBP. The KO orbits are able to maintain their periodicity due to the eigendecomposition of the dynamics, where the system is diagonalized and each eigenfunction is studied in terms of magnitude and shape. Therefore, if desired, unstable contributions from eigenfunctions connected to unstable eigenvalues can be separated.  Figure~\ref{fig:eig} reports the first 24 eigenfunctions from the 3rd order KO solution. The six-dimensional functions are represented with just two dimensions in the $(x,\dot y)$ plane since this information sufficiently describes each Lyapunov orbit. The eigenfunctions are centered at the $\mathcal L_1$ point, meaning that position (0,0,0) corresponds to $\mathcal L_1$ instead of the primary celestial body. Examining a gray-scale gradient, the figure shows the locations of minima in bright white. These curves represent equilibrium conditions, or more precisely, their approximation through a 3rd order KO. Therefore, when we include the initial conditions from the orbits from Figure \ref{fig:L1} in the $(x,\dot y)$ plane as white dots, their locations settle close to the minimum lines. Thus, the Koopman approximation is correctly approximating the dynamics of the system, in terms of its eigenfunctions. However, the accuracy of the approximation decreases as the initial condition becomes larger: the farthest white points are not as well approximated as the white points near the origin. Moreover, the location of the minima is not unique. Many eigenfunctions show well-marked bright lines in the bottom part of the  $(x,\dot y)$ plane, for low velocities. These initial conditions are still connected to periodic orbits. The resulting pathway leaves the Lagrangian $\mathcal L_1$ point and starts orbiting around the main celestial body with a periodic behaviour~\cite{restrepo2018database}. Therefore, even if the eigenfunctions become less accurate the farther they are from the equilibrium point, their minima are still able to connect to different kinds of POs. Furthermore, when looking for periodic orbits, the minima location gives an extremely efficient first guess for the implementation of a  differential corrector that solves for the actual correct periodic initial condition of the KO.   

Note also that Figure~\ref{fig:eig} reports only POs connected to negative $x$ and positive $\dot y$. However, thanks to the symmetric properties of the RTBP, the domain of the eigenfunctions can be widened to support retrograde orbits, and all the remaining POs from axial and central symmetries, as studied and shown by Russel in Ref.~\cite{restrepo2018database}. 

The accuracy of the solutions obtained can be improved by increasing the order of the basis functions at the expense of higher computational costs. The computational burden of the KO solution is a function of the selected order and the dimension of the dynamics of the system, with complexity growing exponentially due to the combinatorial nature of the proposed solution. However, once the Koopman matrix has been calculated and diagonalized, the dynamics of the problem are fully described by the linear combinations of the eigenfunctions. Therefore, the propagation of any state can be performed with ease, making the Koopman theory suitable for spacecraft on-board applications such as navigation (filtering) and attitude control. Most of the computational work can be performed offline, creating a detailed characterization of the dynamics. Any software would just need to have access to the eigenfunctions to evaluate the propagated solution given each starting state, without performing any numerical integration.

%%%%%%%%%%%%%%%%%%%%%%%%%%%%%%%%%%%%%%%%%%%%%%%%%%%%%%%%%%%%%%%%%%%%%%%%%%%%%%%%%%%%%%%%%%%%%%

\section{Halo Orbits Approximation Around $\mathcal L_1$ and $\mathcal L_2$}

\begin{figure}[h!]
	\centering
	\includegraphics[width=\textwidth]{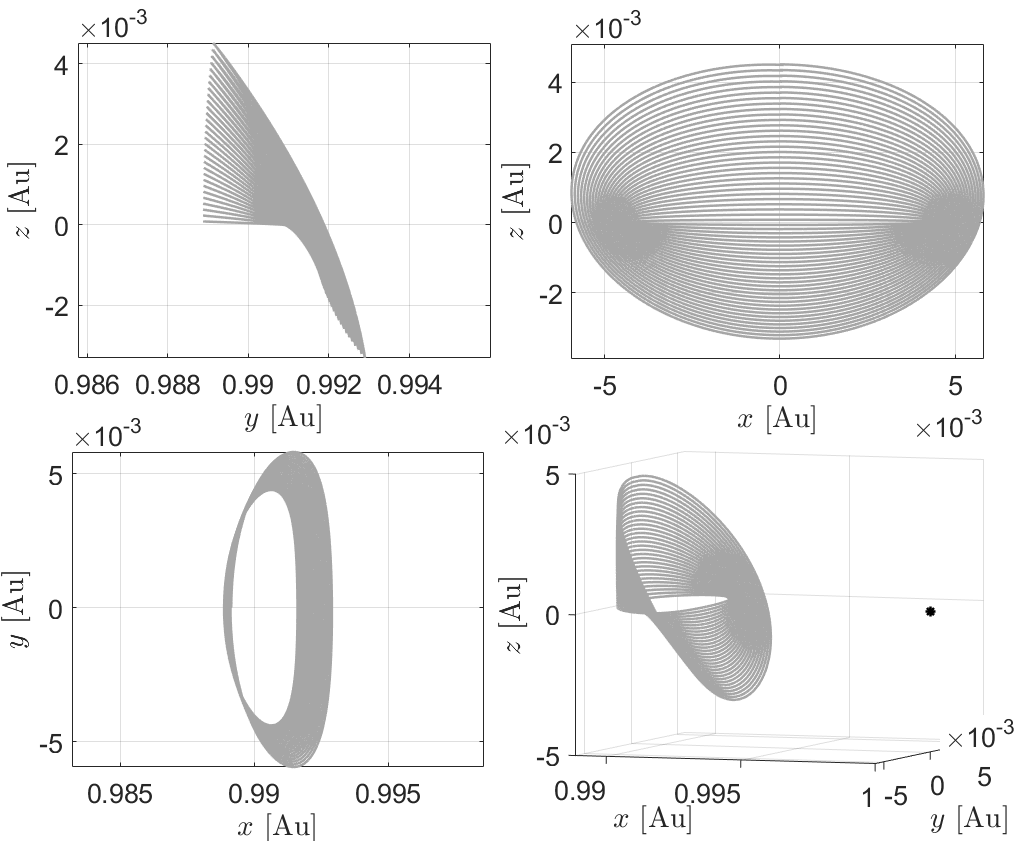}
	\caption{$\mathcal L_1$ Halo orbit family solved by a 5th Order Koopman Operator.    }
	\label{fig:halo}
\end{figure}

The Koopman Operator has also been applied to approximate a second family of POs: the Halo orbits at $\mathcal L_1$ and $\mathcal L_2$. Figure~\ref{fig:halo} reports a set of 30 different orbits calculated through a KO using basis functions of order 5. The figure includes the projection of the orbits in the three reference planes and an isometric three-dimensional view of the trajectories. The black point in the picture represents the secondary celestial body, which is the Earth in the selected Sun-Earth system. As the figure shows, the Koopman solution achieves a robust representation of the orbits. The accuracy of this solution will be later assessed. The periodicity is given by an accurate selection of the initial conditions, which have been found through the implementation of a differential corrector in such a way that the proposed solution describes the same orbits for multiple revolutions. 

\begin{figure}[h!]
	\centering
	\includegraphics[width=\textwidth]{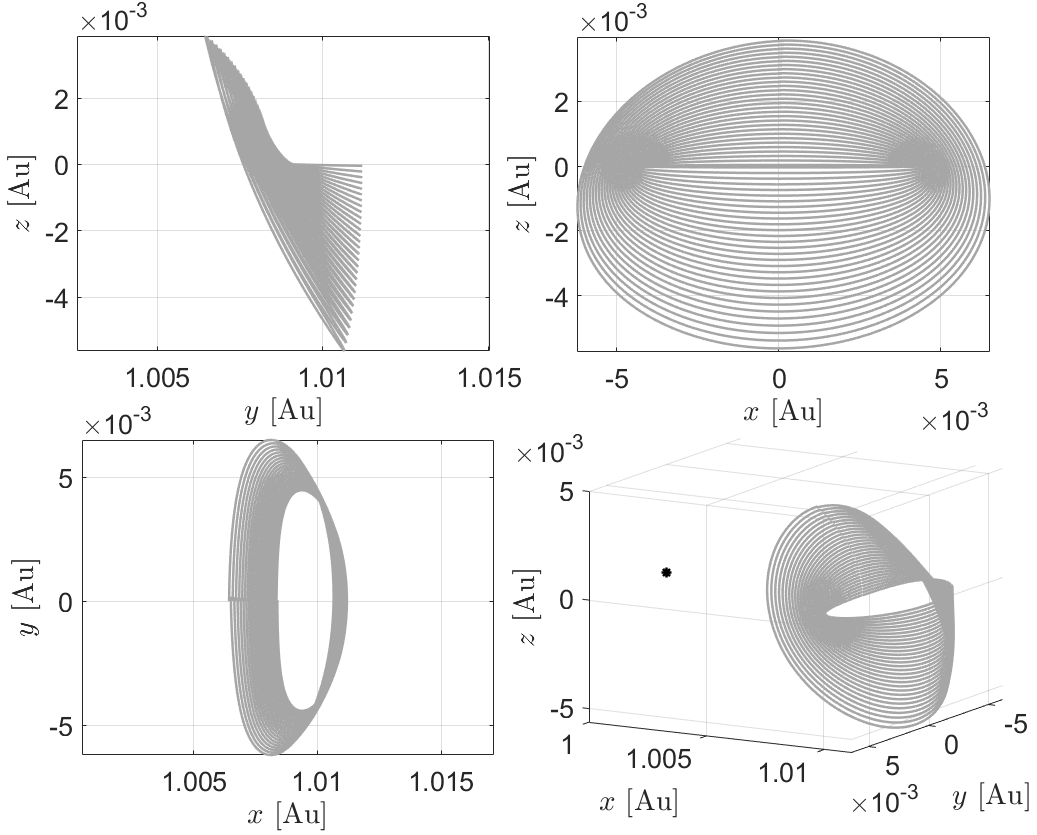}
	\caption{$\mathcal L_2$ Halo orbit family solved by a 5th Order Koopman Operator.    }
	\label{fig:halo2}
\end{figure}

A similar pattern is reported in Figure~\ref{fig:halo2}, showing a family of Halo orbits at $\mathcal L_2$ solved using the 5th order KO. Once again, the black point in the figure represents the secondary body in the RTBP system. The KO is able to accurately represent orbits around both libration points. Since the value of $\gamma$, from the system of Equations~\eqref{eqgamma}, is different when describing the dynamics around $\mathcal L_2$, Richardson's coefficients are different as well, and thus, the computed Koopman matrix is changed for the cases of motion about the $\mathcal L_1$ and $\mathcal  L_2$ points.

\begin{figure}[h!]
	\centering
	\includegraphics[width=0.9\textwidth]{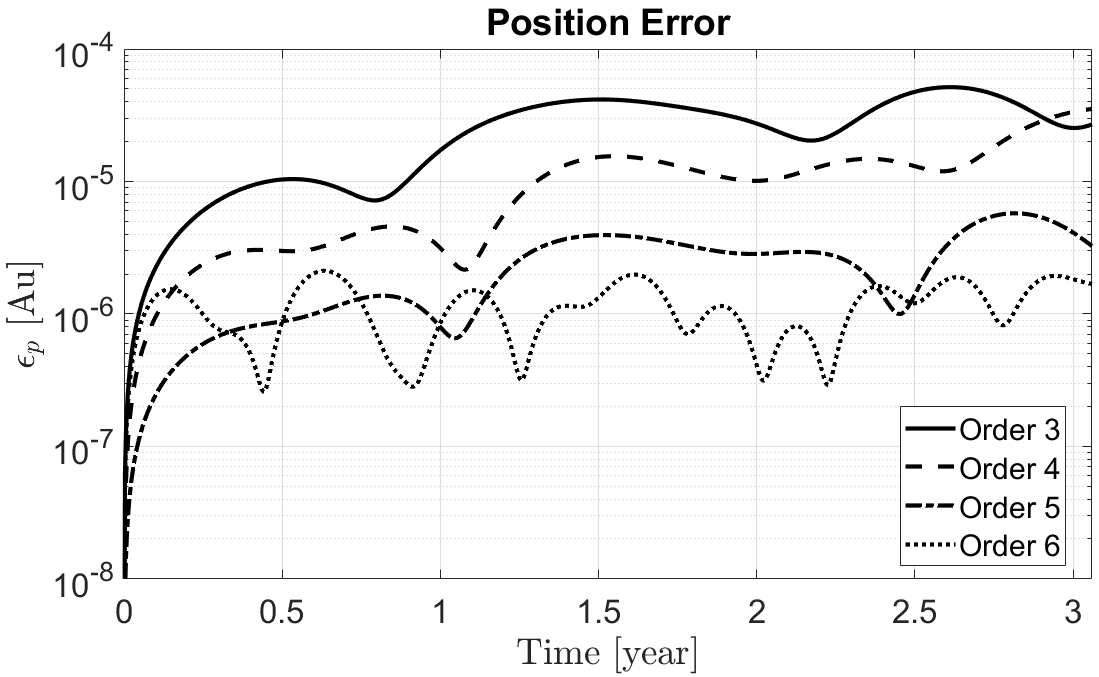}
	\caption{$\mathcal L_1$ Halo orbit accuracy analysis through different KO orders.    }
	\label{fig:halo_order}
\end{figure}

Moreover, convergence analysis of the Koopman operator has been applied by analyzing how the approximation error reduces as the order of the basis functions used increases. Therefore, after selecting a Halo orbit with an initial condition given by:
\begin{equation}
    \bf{ x_0} = 
    \begin{bmatrix}
    0.988882322146701 & 0  & 0.000809201887342 & 0 & 0.008904188320067 & 0
    \end{bmatrix}^T, \label{horbit}
\end{equation}
different orders of the solution have been calculated. Figure~\ref{fig:halo_order} reports the position error, evaluated as in Eq.~\eqref{eq:pos}, for a complete orbit revolution at different orders of basis functions. The continuous line, representing order 3, behaves two orders of magnitude worse than the dotted curve, which is related to the 6th order KO solution. Order 4, dashed line, and order 5, dash-point line, are in between the performance of the other two curves, indicating that the KO accuracy of the solution improves constantly as the number of basis functions becomes larger and thus, we can better represent the non-linearities of the dynamical system. Moreover, while the 3rd order approximation of the Halo solution becomes less accurate as of the propagation proceeds, the 6th order solution maintains the same level of accuracy during the whole orbit, meaning that the second revolution is able to start with accuracy levels comparable to the first one. 

\begin{figure}[h!]
	\centering
	\includegraphics[width=.6\textwidth]{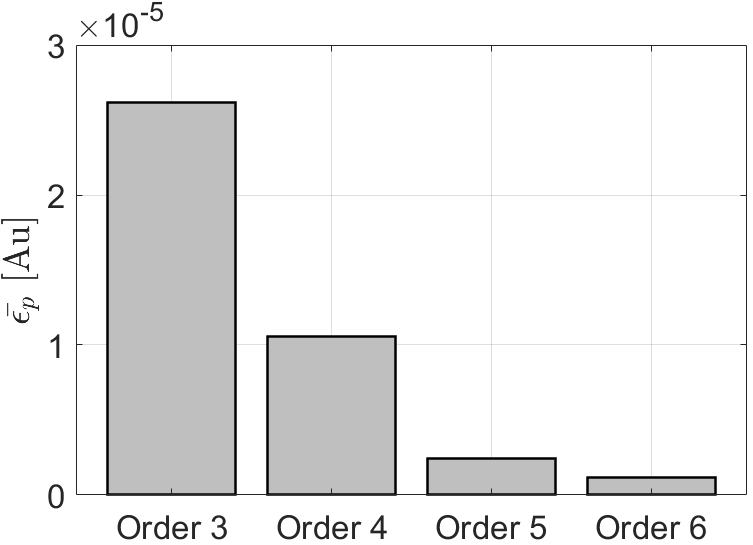}
	\caption{Average position error for different KO orders    }
	\label{fig:bar}
\end{figure}

Figure~\ref{fig:bar} shows the average position error from the KO approximation studied in Figure~\ref{fig:halo_order}. The histogram highlights how the 6th order solution is 22 times more accurate than the 3rd order approximation for the Halo orbit. In the RTBP, the 6th order Koopman Operator approximates the dynamics by using 11 times more basis eigenfunctions than the 3rd order. Therefore, non-linearities are better represented and the performance of the Koopman solution improves.

It is important to note that the benefits of a more precise propagation are paid in terms of computational effort. However, obtaining the KO is a one-time computational effort and thus, once the system has been represented through the eigenfunctions and the observables have been selected, any solution is rapidly evaluated as a simple numerical evaluation of the eigenfunctions at a given time. Therefore, once the basis functions are available and the system has been processed, there is no computational time delay in applying a 3rd or a 6th order solution. 

\begin{figure}[h!]
	\centering
	\includegraphics[width=1.0\textwidth]{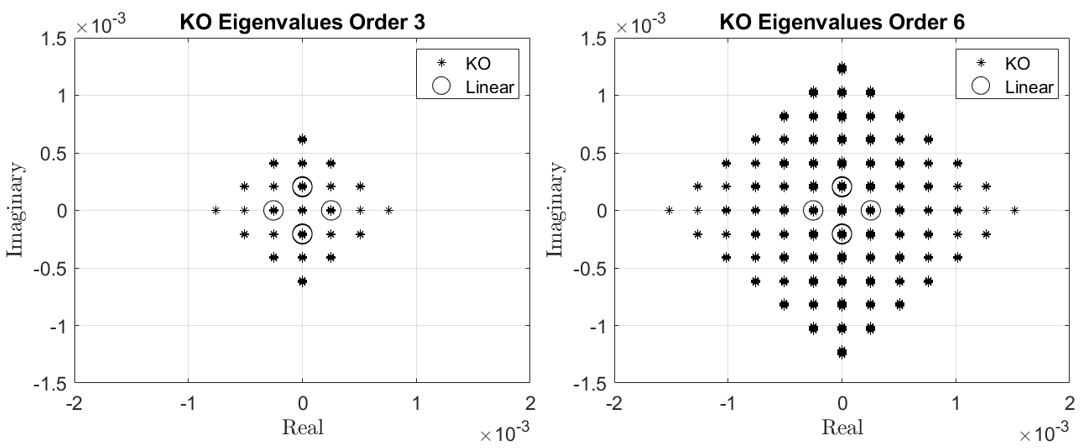}
	\caption{$\mathcal L_1$ Halo orbit accuracy analysis through different KO orders.    }
	\label{fig:rombo}
\end{figure}
 
In addition to the study of the performance of the KO solution, a spectral study can also be done. Figure~\ref{fig:rombo} compares the eigenvalues, computed using the Galerkin method, from the 3rd order and 6th order Koopman matrices. In this particular application, the eigenvalues of the 3rd order solution are a subset of the 6th order approximation. Thus, the 6th order solution incorporates previous orders' information and enhances it with additional nonlinear functions. As such, the location of the high-order eigenvalues is at the edge of the diamond-shaped figure. By looking at the intensity of every single marker, it can be noted how eigenvalues closer to the origin and the imaginary axis are repeated multiple times. Indeed, it is possible to know in advance the position of the unperturbed eigenvalues, for any given order, by the simple composition of the eigenvalues that have already been calculated for lower orders. Therefore, position and repetitiveness are explained in terms of statics and combinatorics (the theory of combinations), where all the new eigenvalues are found as a linear combination of the original ones. However, the eigenvalue composition holds valid only for the linear motion or the unperturbed cases (see Ref.~\cite{koopmanzonal}). Moreover, the linear approximation near the $\mathcal L_1$ is known analytically, as well as its eigenvalues, which have been portrayed in Figure~\ref{fig:rombo} as circles. The linear part of the KO solution matches the linear contribution of the true dynamics. The six linear eigenvalues are perfectly represented and the high-order approximations evolve from the latter to best represent the nonlinear terms of the system of ODEs.

\begin{figure}[h!]
	\centering
	\includegraphics[width=.85\textwidth]{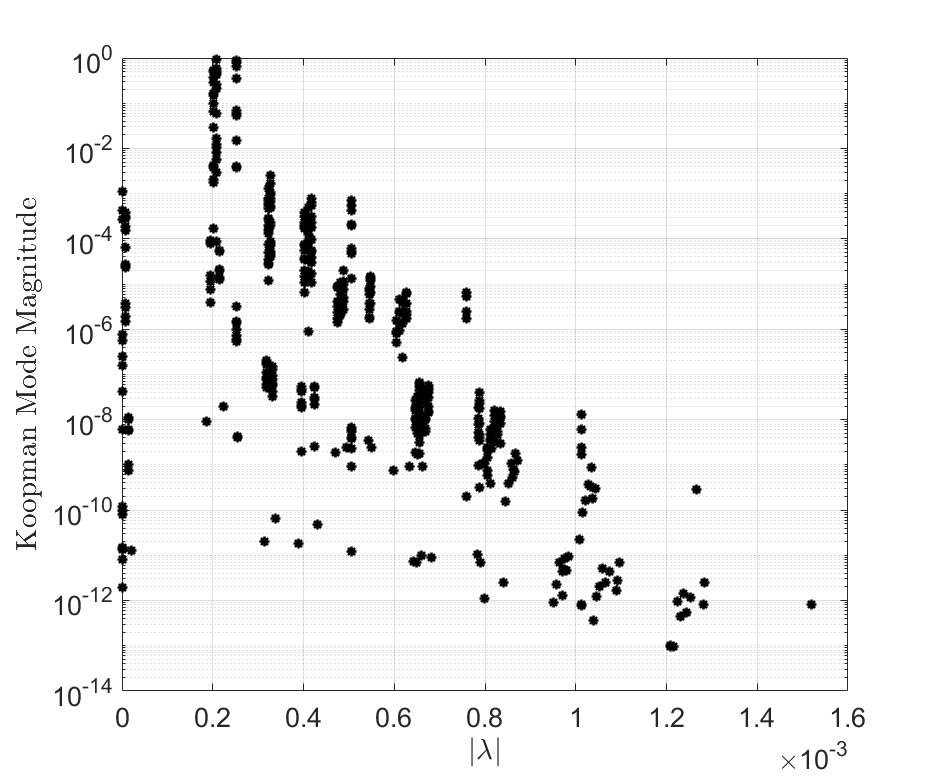}
	\caption{$\mathcal L_1$ Halo orbit mode magnitudes for the 6th Order Koopman Operator.    }
	\label{fig:modes}
\end{figure}

The frequencies and spectral behavior of the system provided by the KO can also be used to assess the convergence of the methodology. This can be done by the study of the Koopman modes, that is, the projection of the observables in the operator. Figure~\ref{fig:modes} presents the intensity of the Koopman modes of the system. Particularly, it shows how the eigenvalues with the highest module are the least influential in the solution. However, the small contributions from the modes connected to large eigenvalues are crucial to achieving an accurate approximation of the dynamics, and they make the difference in accuracy when comparing the order 3 solution with the one of order 6. Therefore, many terms other than the linear ones are non-zero, with influence that decreases as the order of the eigenfunctions increases. This also implies that the Koopman modes decay with larger frequency values, leading to a possible truncation of the expansion for eigenfunctions with, particularly high frequencies. 

\begin{figure}[h!]
	\centering
	\includegraphics[width=1.0\textwidth]{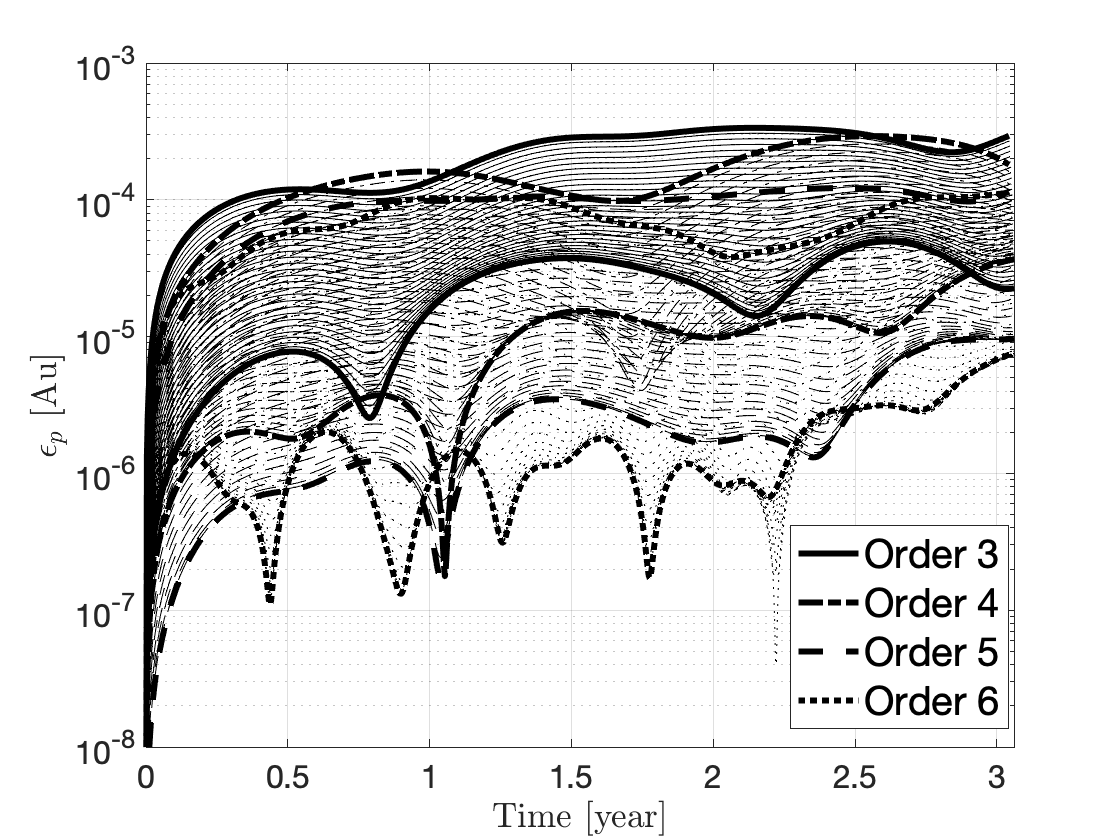}
	\caption{Convergence analysis of the Halo orbit solution of the KO for different orders.    }
	\label{fig:cool}
\end{figure}

A similar convergence analysis can be done on the Halo POs represented in Fig.~\ref{fig:halo}. Particularly, for each PO, the position error has been evaluated by comparing the Koopman solution with a classical Runge-Kutta numerical integration. Figure~\ref{fig:cool} shows, with different line styles, the accuracy of the approximation on each orbit for different orders of the KO. As expected, as the Halo orbits become larger, the position error increases consequently. Therefore, for any given order, the position error convergence analysis describes a set of curves. In that regard, order 3, represented by a continuous line, is the least accurate solution, while order 6 is the most accurate. However, it is worth noticing how, for each orbit, the gain in accuracy obtained by increasing the KO order behaves differently depending on the initial condition. The KO order 6 solution is two orders of magnitude more accurate than the 3rd order for orbits close to $\mathcal L_1$. However, as the initial condition of the Halo orbit moves away from the equilibrium point, the gain in accuracy is reduced, settling around a single order of magnitude. Indeed, for very large initial conditions, increasing the order of the basis functions to over a given value stops being beneficial since the gain in performance does not repay the higher computational burden. This aspect is connected with the Richardson approximation applied for the polynomial representation of the dynamics. 

\begin{figure}[h!]
	\centering
	\includegraphics[width=0.9\textwidth]{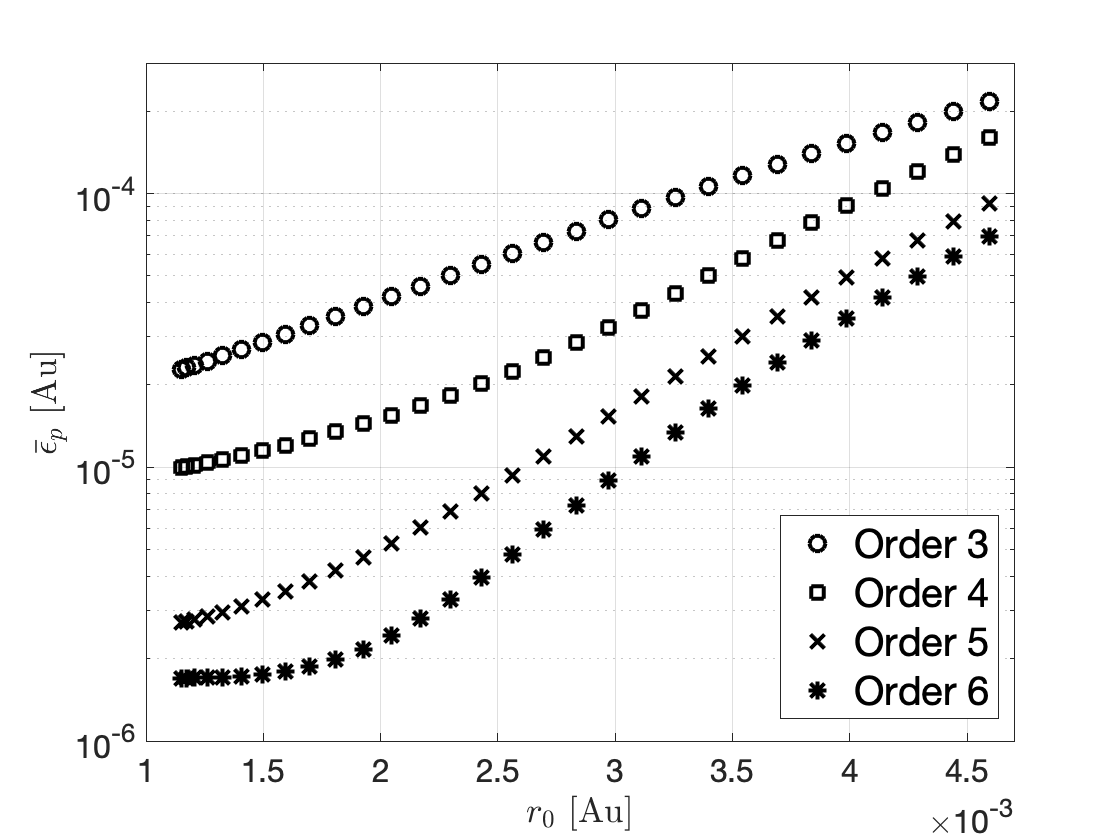}
	\caption{$\mathcal L_1$ Halo orbits convergence and accuracy analysis for different KO orders.  }
	\label{fig:acc}
\end{figure}

The Richardson's ODEs representation fails to provide an accurate approximation of the system for initial conditions far from the equilibrium point. Consequently, looking back at Figure~\ref{fig:halo}, it can be noted that for each set of curves, at different orders, there is a well-marked separation in the bottom part of the graph, while curves from different orders start to get closer as the orbit becomes larger. This aspect has been highlighted in Figure~\ref{fig:acc}, where the mean position error ($\bar{\epsilon}_p$) for each orbit is compared with the size of the Halo orbit ($r_0$) and the maximum order of the basis functions used to represent the KO. Thus, orbits with the furthest initial conditions from the $\mathcal L_1$ point generate mean position errors that range in value by up to one order of magnitude. However, as we get closer to $\mathcal L_1$, and thus reducing $r_0$, the mean position error for each KO solution order considered decreases with a different slope, making the accuracy gain more substantial as we increase the order of basis functions used. Additionally, it can also be observed that there is a region very close to $\mathcal L_1$ where the error seems to reach convergence and stabilize for 6th order basis functions. Thus, for Halo orbits close to $\mathcal L_1$, an increase of the KO order leads to a substantial improvement in accuracy and in the performance of the approximation of the system. 

\section{Comparison with Extended Dynamic Mode Decomposition}

We have seen before that the Galerkin method can be used to approximate the Koopman Operator in closed-form. In this section, we obtain the Koopman Operator using Extended Dynamic Mode Decomposition (EDMD), which is a numerical technique that has been extensively used in non-linear engineering problems in the last decade. The aim of the comparison is to assess that the Galerkin methodology is a far superior technique compared to EDMD both in terms of accuracy and robustness of the approximation. Consequently, this chapter shows that a complete EDMD approximation analysis is not needed because of its approximation errors and the performance of its solution when compared to the Galerkin method. EDMD is related to the KO that we presented in previous sections and provides a relatively simple tool for solving nonlinear KO models. This section describes the EDMD algorithm at a high level; for more details please refer to the work by Williams~\cite{williams2015data}. 

The EDMD approach attempts to estimate a finite-dimensional representation of the Koopman operator $\mathcal{K}$ using a linear mapping of the Koopman modes.  The EDMD approach is restricted to discrete time systems in the form:
\begin{equation}
{\bf x}_{k+1}={\bf F}({\bf x}_k),
 \end{equation}
where ${\bf x}_k$ are the discrete states and ${\bf F}$ is the nonlinear dynamic model. 
In the same fashion as in the Galerkin method, eigenvalue decomposition is used to
find the $\mathcal{K}$ that minimizes a loss function related to the EDMD prediction error over the data samples (See Ref.~\cite{williams2015data} for full description). The EDMD solution is given by:
\begin{equation}\label{EDMD_solution}
\mathcal{K} = G^\dag A
\end{equation}
where $\dag $ denotes the pseudoinverse, $G$ and $A$ are defined as:
\begin{eqnarray}
    G & = & \frac{1}{M}\sum_{k=0}^{M-1}\Psi^T({\bf x}_k)\Psi({\bf x}_k), \nonumber \\
    A & = & \frac{1}{M}\sum_{k=0}^{M-1}\Psi^T({\bf x}_k)\Psi({\bf x}_{k+1}),
\end{eqnarray}
and $\mathcal{K} ,G,A\in \mathbb{C}^{K \times K}$. It is important to note that the solution for $\mathcal{K}$  given above 
is a finite dimensional approximation of the KO. Furthermore, with the finite dimensional approximation of the KO given by Eq.~\eqref{EDMD_solution}, we can 
determine an approximation of the eigenfunctions and eigenvalues of KO. 

\begin{figure}[!ht]
 \begin{center}
        \psfrag{x1}[][]{\footnotesize{$x_1$}}
        \psfrag{x2}[][]{\footnotesize{$x_2$}}
        \psfrag{Time (min)}[][]{\footnotesize{Time (min)}}
        \includegraphics[keepaspectratio,
        width=0.65\textwidth]{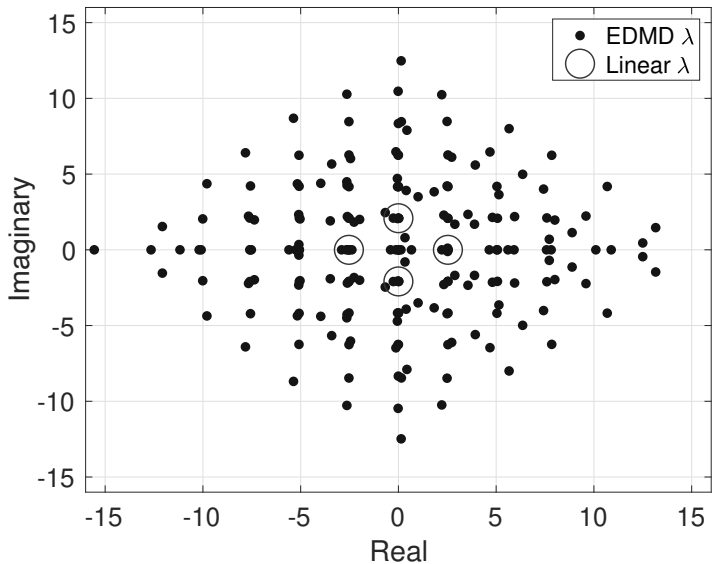}\label{EDMD_eig}
        \caption{\bf  Extend Dynamic Model Decomposition}\label{EDMD}
 \end{center}
\end{figure}

For the EDMD to work, we need to provide some numerical information about the dynamical system. To that end, initial conditions are drawn from a normal distribution with ${\bf x}\sim\mathcal{N}\left({\bf 0},\sigma^2 I_{4\times4}\right)$, where $\sigma$ is the scale used in the Gaussian weighting function when computing the Galerkin inner product. Note that due to its numerical nature, this method is not as accurate as the Galerkin approximation. Figure~\ref{EDMD} shows the eigenvalues computed for the Richardson third-order model using the EDMD approach. The linear approximation near $\mathcal L_1$ is known to have two real eigenvalues and four complex eigenvalues (in six dimensions). These linear eigenvalues, which define the shape of the spectrum, can be found analytically and they are matched by the EDMD. However, EDMD shows distortions on the higher-order eigenvalues that result from the linear ones, where the Galerkin method does not have this distortion. The predicted accuracy for the EDMD method is not shown since the error is extremely large and the EDMD solution does not provide a meaningful approximation of the KO for the three-body problem. The authors conclude that the three-body problem is poorly approximated by the EDMD method because this method relies on numerically approximating the KO and this approximation introduces too much error (in the eigenvalues and eigenvectors of the KO). In particular, the unstable eigenvalues of the RTBP make it challenging to numerically approximate the KO matrix form.  However, using the closed-form computation of the KO avoids this approximation error and results in a far superior solution.

%%%%%%%%%%%%%%%%%%%%%%%%%%%%%%%%%%%%%%%%%%%%%%%%%%%%%%%%%%%%%%%%%%%%%%%%%%%%%%%%%%%%%%%%%%%%%%
% Orbit Keeping
%%%%%%%%%%%%%%%%%%%%%%%%%%%%%%%%%%%%%%%%%%%%%%%%%%%%%%%%%%%%%%%%%%%%%%%%%%%%%%%%%%%%%%%%%%%%%%
\section{Closed-Loop Station-Keeping for a Halo Orbit}

In previous sections, we show that the Koopman solution leads to accurate approximations of the dynamics of the RTBP while providing a linearized system of these dynamics. As such, these properties can now be used to generate an orbit-keeping application, where the goal of the approach is to keep a spacecraft orbiting in a specific Halo orbit. 

We first assume that for the initial condition, the spacecraft has an offset with respect to the desired pathway. In addition, we consider the case where the state of the satellite is provided by some measurements from sensors or the ground stations. Therefore, the actual position and velocity vectors of the satellite need to be estimated and based on that, the spacecraft has to be redirected and controlled to achieve the desired behavior. Consequently, an estimation and control system based on the KO is proposed. Given GPS measurements from the true trajectory, the state of the spacecraft, in terms of its position and velocity, is estimated through a KO version of the particle filter. Afterward, reaching and keeping the desired orbit is achieved through a nonlinear model predictive controller (NMPC), which uses the KO solution for the integrative part of the control. 

\subsection{The Koopman Operator Particle Filter - KOPF}

The particle filter (PF) is a well-known nonlinear estimator that predicts the state of a system by approximating probability density functions (PDFs) through a set of samples~\cite{ristic2003beyond}. The PF is a robust and accurate filter. However, its use in astrodynamics applications is limited by the high computational burden requested by the time propagation of every single particle, an aspect that makes the filter unsuitable for onboard applications. After initializing an ensemble of particles according to the given initial state PDF, each sample is integrated forward in time until a measurement becomes available. Thus, each particle is propagated numerically according to the selected numerical integrator. Once measurements are given, each particle receives an importance weight which is proportional to the likelihood given the measurements. After normalizing the weights such that they sum the unity, the estimated mean and covariance are evaluated as a weighted mean among all of the particles. Before starting the next step and integrating the ensemble forward in time, particle resampling is performed in order to keep the system tractable. The particles are therefore resampled according to their own weight, as in the Bootstrap Particle Filter (BPF)~\cite{douc2005comparison}, or using a Gaussian approximation of the \textit{a posteriori} PDF, given the posterior mean and covariance, as in the Gaussian Particle Filter (GPF) ~\cite{kotecha2003gaussian}.       

The main drawback of any particle-based filtering technique is performing multiple integrations, which are time-consuming and drastically increase the machine workload. Thus, different techniques where the ensemble of particles of Monte Carlo filters are evaluated through a polynomial at a given time step have been developed~\cite{Servadio2021}.  Therefore, a new particle filter based on the KO is proposed, where each integration is substituted by the mere evaluations of functions using the KO approximation of the solution for the dynamics. In particular, the Koopman Operator Particle Filter (KOPF) evaluates the propagated particles, each time step, by considering their different initial condition during the evaluation of the eigenfunctions. After the dynamics have been analyzed and linearized in the set of basis functions, the state of each particle, for any given time, is a function of its own initial conditions. Looking back at Eq.~\eqref{solution}, each particle is propagated through the evaluation of different ${\bf x}(t_0)$ into the eigenfunctions. Therefore, all the particles are integrated in the KO framework and the state PDF is propagated rapidly, making the KOPF suitable for on-board applications. The measurement update part of the filtering algorithm remains untouched and it continues normally according to the BPF or GPF mathematics, and following the selected resampling technique.

\subsection{The Nonlinear Model Predictive Controller - NMPC}

The Koopman theory is also integrated into the control segment of the station-keeping application. The KO solution is therefore merged inside a model predictive controller, where the control feedback input is optimized by predicting the future behavior of the state of the system given the current state.

The main idea of the Nonlinear Model Predictive Controller, NMPC, is that, at each sampling time step, the predicted future behavior of the system is optimized over a finite time horizon~\cite{grune2017nonlinear}. The optimal control sequence is then used as feedback control values to modify the state of the spacecraft at the current time step. The prediction of the future behavior of the state is compared to a given constant reference. The cost function to be used in the optimization should penalize the distance of the estimated state from the filter with respect to the reference trajectory. As expected, increasing the prediction horizon length directly relates to a more accurate feedback control input, which will lead the NMPC to achieve better performance. However, it is often convenient to penalize the control input $\bf u$ for modeling purposes, such as avoiding control values that are connected to high energy. For station-keeping applications, this aspect connects to fuel optimization in the particular case that the selected cost function analyzes the amount of fuel requested by the spacecraft. Constraints, both on the control, as well as on the states, can be easily taken into account for the classic NMPC algorithm~\cite{grune2017nonlinear}. The basic NMPC algorithm can be summarized in three major steps. At first, the state of the system is measured. This step is achieved through the state estimation performed by the KOPF. Then, an optimal control problem (OCP) is solved, with the goal of minimizing some well-defined cost function with respect to the control input $\bf u$, and subject to either control or state constraints. Lastly, the NMPC feedback value is selected and used as a control value for the next sampling period. Open-loop optimal control is also possible~\cite{patwardhan1990nonlinear}. 

The Koopman operator is used inside the NMPC for the prediction horizon length, where the state of the system is propagated forward in time to assess the validity of the proposed control input value. Therefore, the newly proposed KO solution of the RTBP is embedded both in the estimator and in the controller part of the closed-loop station-keeping algorithm.

\subsection{Numerical Results}

The closed-loop station-keeping technique described before has been applied to the control of a spacecraft about a Halo orbit defined by Eq. (\ref{horbit}). After the coordinate transformation presented in Eq.~\eqref{coord}, which translates the origin of the system to the $\mathcal L_1$ point and scales the problem by the distance between the secondary body and $\mathcal L_1$ itself, the new initial conditions are:
\begin{align}
    \bf{x}_{0,T}& = 
    \left[
    -0.084775484328489 \quad 0.03  \quad 0.111161030000000 \quad 0.003 \quad 0.896068969160230 \quad 0.003
    \right]^T,  \\
    \bf{x}_{0,D} &= 
     \left[
    -0.114775484328489 \quad 0  \quad 0.081161030000000 \quad 0 \quad 0.893068969160230 \quad 0
    \right]^T,
\end{align}
where $\bf{x}_{0,T}$ indicates the initial true position of the spacecraft. Note that this represents a large offset with respect to the desired initial position $\bf{x}_{0,D}$. The initial estimate is then selected randomly from a Gaussian PDF, with the true initial state as the mean value and a covariance matrix given by:
\begin{equation}
\bf{P}_{0} = 
    \begin{bmatrix}
    10^{-6}\bf{I}_{3x3} & \bf{0}_{3x3} \\
    \bf{0}_{3x3} & 10^{-7}\bf{I}_{3x3}
    \end{bmatrix},
\end{equation}
where $\bf{I}_{3x3}$ indicates the identity matrix, and $\bf{0}_{3x3}$ the null matrix. The satellite is controlled such that the maneuver corrections and observations are performed every 4 days. Additionally, it is assumed that the thruster uses a negligible amount of fuel compared to the satellite's mass, meaning that the mass of the spacecraft remains constant during the simulation. Finally, the system is considered to be subject to Gaussian process noise $\mu \sim \mathcal{N}(\bf{0}_{6\times 1},10^{-5}\bf{I}_{6 \times 6})$, and the measurements are obtained considering that the spacecraft is equipped with GPS with a Gaussian observation noise $\nu \sim \mathcal{N}(\bf{0}_{3 \times 1},10^{-5}\bf{I}_{3 \times 3})$.

The NMPC optimizes the control input following the cost function $J = {\bf u^T}W{\bf u}$, where $W$ is a weighting matrix, selected as the identity in the current application. Constraints are given for the maximum feedback control input that the thruster can generate in the NMPC: a value of $\pm0.00474$ km/s is selected. 

\begin{figure}[h!]
	\centering
	\includegraphics[width=.65\textwidth]{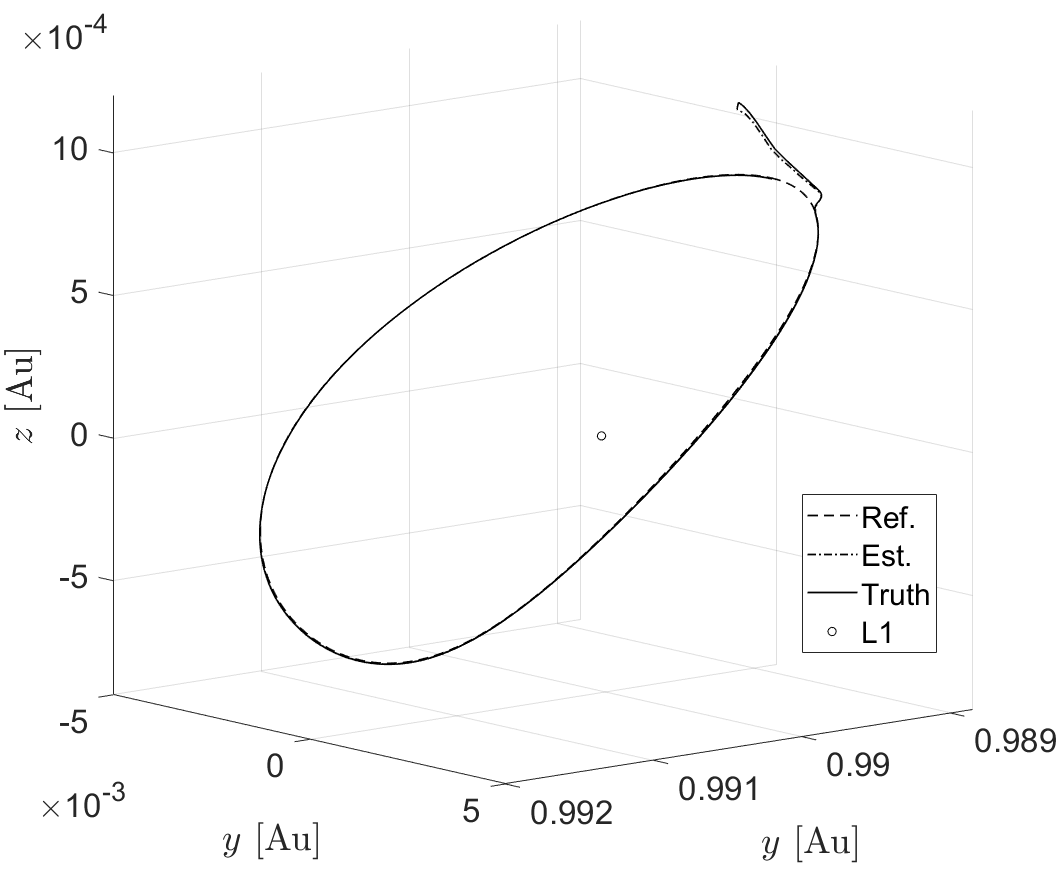}
	\caption{Station-keeping application of a spacecraft orbiting in a Halo orbit around $\mathcal L_1$.   }
	\label{fig:keep}
\end{figure}

The station-keeping results using a KO with basis functions of order 3 are presented in Figure~\ref{fig:keep}, where the reference orbit, the true position of the satellite, and its estimated position, are shown respectively as dashed, continuous, and point-dashed lines. As can be observed from the figure, the KOPF and NMPC are able to control the spacecraft and bring it to the desired pathway after just a few time steps. Particularly, the initial offset can be noted at the beginning of the simulation, and soon after, the spacecraft is already in the trajectory of the Halo orbit. In this example, the NMPC prediction horizon takes into account the following 10-time steps, which avoid the overshooting during the approaching to the desired orbit, and provides a smooth approach to the desired state. This means that after the quick transient, the three lines overlap for the remaining part of the trajectory, assessing the validity of the proposed technique. The control horizon is set for 3 time steps.

\begin{figure}[h!]
	\centering
	\includegraphics[width=.7\textwidth]{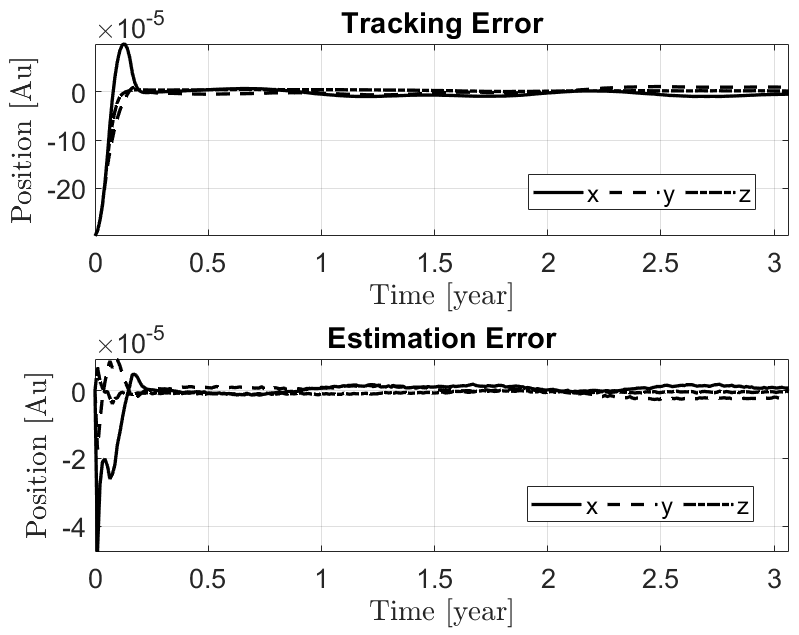}
	\caption{Tracking and estimation error of a spacecraft orbiting in a Halo orbit around $\mathcal L_1$.   }
	\label{fig:err}
\end{figure}

In this regard, it is also possible to assess the tracking error of the spacecraft. This is done by evaluating the difference between the reference orbit and the true position. Analogously, the estimation error is calculated as the difference between the true position of the satellite and the estimated state from the filter. Figure~\ref{fig:err} shows the tracking and the estimation errors for the presented application. After a quick transient where the system corrects the initial offset, the error levels settles down and it shows an accurate and robust estimate and control of the dynamics of the spacecraft.

\begin{figure}[h!]
	\centering
	\includegraphics[width=.7\textwidth]{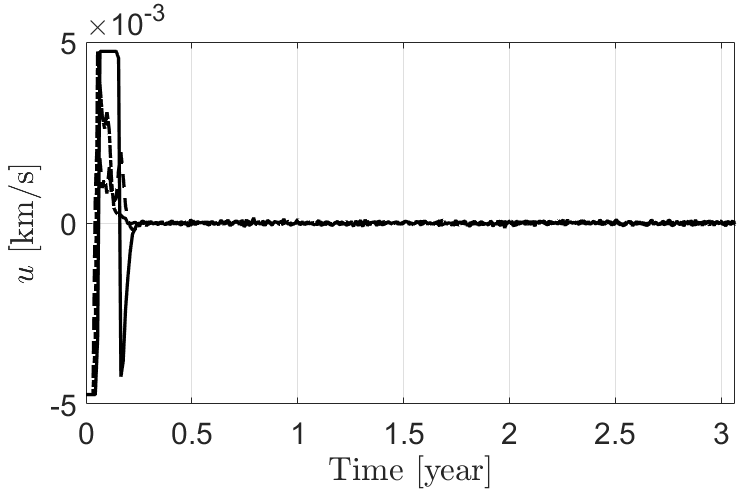}
	\caption{Feedback control input for a spacecraft orbiting in a Halo orbit around $\mathcal L_1$.   }
	\label{fig:uu}
\end{figure}

The duration of the transient is directly connected to the maximum thrust available from the thrusters. Thus, in the first time steps, while compensating for the initial offset, the thrusters are saturated and reach the maximum available thrust. Having a more powerful thruster would reduce the transient time and achieve a quicker match between reference and target orbits. Figure~\ref{fig:uu} shows the feedback control input, in terms of change in the spacecraft's velocity, during the complete simulation.

%%%%%%%%%%%%%%%%%%%%%%%%%%%%%%%%%%%%%%%%%%%%%%%%%%%%%%%%%%%%%%%%%%%%%%%%%%%%%%%%%%%%%%%%%%%%%%
% Conclusions
%%%%%%%%%%%%%%%%%%%%%%%%%%%%%%%%%%%%%%%%%%%%%%%%%%%%%%%%%%%%%%%%%%%%%%%%%%%%%%%%%%%%%%%%%%%%%%

\section{Conclusions}

The Koopman Operator methodology presented in this work is shown to provide an accurate transformation that embeds the nonlinear dynamics in a global linear representation. The selected application, the circular restricted three-body problem, is a highly nonlinear and highly chaotic system, whose periodic solution is hard to obtain. Nevertheless, the Koopman approximation provides an accurate analytical representation of the dynamics of the system, being able to accurately reproduce both Lyapunov and Halo orbits as a linear combination of well-selected eigenfunctions. As such, the dynamics of the system are transformed into a new framework, where each contribution has a linear dependence on the overall solution. 

An analysis of the accuracy of the KO solution supports the validity of the newly proposed technique, especially for those initial conditions close to the equilibrium points, as shown by the study of the location of the minimum curves in the eigenfunctions.  
In addition, the accuracy of the newly proposed techniques has been assessed through a convergence analysis, where the position error of different families of Halo orbits has been compared to classical numerical integration techniques for the different orders of the KO. The results show that the behavior of the KO framework is similar to the perturbation theory since the robustness of the KO approximation decreases the further the initial conditions are selected from the equilibrium points.

Moreover, the KO mathematics has been proven helpful to provide the spectral behavior of the system. Particularly, a study of the frequencies and modes of the dynamics is included with this work. As such, periodic orbits are found by studying the eigenfunctions of the system directly and their curves of minima. 

Furthermore, the KO has been applied to control problems. In particular, a station-keeping application is proposed trhough a new filtering technique and a model predictive controller developed in the KO framework. The KOPF shows high accuracy levels for spacecraft state estimation, while the NMPC uses the KO solution to optimize the control feedback input for the satellite. 

Lastly, a quick comparison of different methodologies for the evaluation of the Koopman matrix is presented. The analytical approach proposed in this paper is shown to better describe the spectral behavior of the dynamics of the system when compared to data-based methods such as the EDMD.

In conclusion, the KO provides a novel alternative to classical numerical integration techniques for solving  Ordinary Differential Equations. The new methodology provides, at the same time, the ODE's solution and a deep understanding of its characteristics, in terms of stability and robustness, achieved thanks to the knowledge of the system eigenvalues and modes. Therefore, the new framework holds promise for a variety of future applications including onboard navigation applications (estimation and/or control), filtering problems, and Attitude Determination and Control Systems~\cite{chen2020koopman}.   

%%%%%%%%%%%%%%%%%%%%%%%%%%%%%%%%%%%%%%%%%%%%%%%%%%%%%%%%%%%%%%%%%%%%%%%%%%%%%%%%%%%%%%%%%%%%%%
%%%%%%%%%%%%%%%%%%%%%%%%%%%%%%%%%%%%%%%%%%%%%%%%%%%%%%%%%%%%%%%%%%%%%%%%%%%%%%%%%%%%%%%%%%%%%%
%%%%%%%%%%%%%%%%%%%%%%%%%%%%%%%%%%%%%%%%%%%%%%%%%%%%%%%%%%%%%%%%%%%%%%%%%%%%%%%%%%%%%%%%%%%%%%

\section*{Acknowledgments}

The authors wish to thank Hailee Hettrick at Massachusetts Institute of Technology for her help in reviewing and improving this manuscript. The authors want to acknowledge the support of this work by the Air Force’s Office of Scientific Research under Contract Number FA9550-18-1-0115.

\bibliography{libration_koopman.bib}

\end{document}